FULL FACT

# AI-Generated Disinformation in the UK: Risk of Harm, Context, and Classification

## Full Fact / University of Westminster: AISI Challenge Fund

Peter Cunliffe-Jones[1], Joseph O'Leary[2], David Corney[2] and Andrew Dudfield[2]

[1] P.Cunliffejones@westminster.ac.uk University of Westminster, UK
[2] joseph@fullfact.org, david.corney@fullfact.org, andrew.dudfield@fullfact.org
Full Fact, London, UK

*This work was supported by the AI Security Institute (ASI) via the AISI Challenge Fund*

# Executive Summary

**More people across the UK are today exposed to more AI-generated and AI-altered disinformation and misinformation than ever before** - exposing individuals and society to a growing range of harmful consequences and potential consequences.

Analysis of a sample of different types of AI-generated videos, fabricated audio, manipulated images and false statements from chatbots - examined for this report - found the sample had a substantive risk of causing or contributing to harm to individuals and society in eight distinct fields, from contributing to incidents of serious social unrest and vigilante violence to causing direct harms to health and causing the sort of serious financial loss that can be caused by online scams and fraud. Other fields of risk included: abuse serious enough to affect individuals' health and behaviour; public engagement with the police and justice systems; susceptibility to false conspiracy theories with the potential to cause direct harms; and broader changes to social and political attitudes with potential to affect political, social events over the longer term.

This report – produced using a harm-risk assessment model developed in a four-year research project published by University of Westminster Press in 2025[1] – analyses evidence from a dataset of 112 pieces of AI-generated or AI-altered disinformation or misinformation seen tens of millions of times across the UK between 1 January 2025 and 31 March 2026. The dataset of examples studied does not, of course, provide an exhaustive sample of all AI-generated false information in circulation in the period, but rather a snapshot of examples showing some, but not all, of the potential effects.

From this dataset, our analysis identified 94 of the 112 entries (83.9%) as creating a substantively false or misleading understanding, i.e. a broad disinformation effect. The remainder - 18 of the entries (16.1%) - were identified as creating an understanding that was only narrowly inaccurate, and thus no such disinformation effect. This means that more than four in five pieces of content we assessed added to reasons for the public to distrust information as not merely inaccurate but substantively false or misleading: a broad disinformation effect with potential for significant effects for society.

Beyond this broad disinformation effect, our analysis identified

[1] See: (i) Peter Cunliffe-Jones, "Fake news: What's the harm?" 2025, *University of Westminster Press*. https://doi.org/10.16997/mpub.14614695; (ii) Peter Cunliffe-Jones, "Fact checking what matters: How a harms-based model for selecting claims works", *Harvard Kennedy School (HKS) Misinformation Review,* July 2026. https://doi.org/10.37016/mr-2020-197.

- 46 of the 112 entries (41.1%) as having a substantive potential to cause or have caused or contributed to, one or more of the specific substantive consequences set out in the model
- 66 of the entries (58.9%) were found to have either no or very limited potential for consequences.

## Potential for societal consequences identified in 8 broad fields

The harm-risk model, explained below, identified the false claims in the dataset as having potential - either to cause harm as a direct effect of public exposure to that specific piece of false information, or by contributing to the cumulative effect of repeated exposure to that or similar claims - with harms in eight broad fields:

(1) Incidents of serious social unrest and/or individual vigilante violence;
(2) Incidents of abuse sufficiently serious to affect individuals' health, behaviour;
(3) Direct financial harm to individuals via online scams, hoaxes;
(4) Harm to individual and public health;
(5) Public engagement with the police, justice system;
(6) Public attitudes to climate issues that impact public policy;
(7) Susceptibility to false conspiracy theories with potential to cause harms;
(8) Public social and political attitudes that influence political, social events.

Our analysis found that the potential outcomes could affect different sizes of groups. In one case, a specific individual could have been targeted by abuse sufficient to affect health or behaviour. Eight of the outcomes could have affected individuals not specific to a particular group - for example scams that cause a financial loss and false claims about health treatments that harm the health of individuals. Ten of the outcomes could have affected members of a specific group or community - for example abuse of Jews, Muslims, members of the police or trans people. And 39 of the outcomes could have affected a large part, or the whole of society, for example via incidents of social unrest, effects on political or social attitudes or effects on climate mitigation policies.

Given the role that disinformation and misinformation has played in events in the United Kingdom in recent years – from financial loss caused to individuals by online scams,[2] to public health during the Covid-19 Pandemic,[3] to the post-Southport riots[4] – and the context facing the United Kingdom in the coming years, we believe these risks to be serious and growing.

[2] Online Information Advisory Committee "Understanding Online Financial Harm Fraud, Scams and Disinformation Exposure Across Demographic Groups in the UK", *Ofcom*, 27 November 2025 https://www.ofcom.org.uk/siteassets/resources/documents/about-ofcom/structure-and-leadership/online-information-advisory-committee/understanding-online-financial-harm-fraud.pdf?v=408231

## Recognising much false information is inconsequential

At the same time, it is important to recognise that, while all entries were in some way false, a clear majority - 66 of 112 entries - were found to have either no or very limited potential to cause or contribute to specific, substantive consequences, or harms.
In 17 cases, this was due to the claim being only narrowly inaccurate and hence having no potential for a misinformation or disinformation effect. In 19 cases, this was due to the topic itself being inconsequential. In 24 cases, this was because evidence showed that insufficient people believed the claims to cause the potential effects. And in the remainder of cases evidence showed that those who believed the claim did not have both the capacity and motivation to act on the false understanding it caused.

The potential of some AI-generated disinformation and misinformation to cause or contribute to serious harm, and for other such information to have limited potential for impact, reinforces the need for policy to distinguish the potentially harmful and the inconsequential.

## Common news themes, impersonation & fabrication: features of AI-gen disinformation

The most frequent topic in the dataset related to what could be broadly described as "politics", a theme appearing in 71 (63.4%) of the 112 dataset entries. Other common themes our analysis identified were: events related to foreign countries; changes to the individual economy; the activities of the police or criminal justice system; the wars in the Middle East; the actions of particular ethnic or religious communities; well-known conspiracy theories and claims related to health and medicine. With the exception of some of the conspiracy theory content, a strong correlation could be seen between themes in the dataset and events, people or situations appearing in mainstream news.

[3] See this study "The impact of misinformation on the COVID-19 pandemic" https://pmc.ncbi.nlm.nih.gov/articles/PMC9114791/
[4] See this DSIT committee report: https://committees.parliament.uk/work/8641/social-media-misinformation-and-harmful-algorithms/publications/.

Within the topic of "politics" the commonest format was AI-generated video, or video and audio purporting to show prominent politicians or establishment figures announcing political actions or policies that were not real. The other main format was fabricated content purporting to show situations or events that had not, in fact, occurred.

As generative AI tools have improved and become more widely used, the previous barriers to creating misleading content like this, at scale, have been weakened. Obvious glitches such as inaccurate limbs or body parts on people are less frequent today than they were. And with low levels of media literacy, existing defence mechanisms to this sort of information threats are limited. New forms of detection tools such as the SynthID invisible watermark system from Google can assist expert users to identify inauthentic video. Recently, Google announced partnerships with Nvidia and OpenAI to introduce genAI watermarks across multiple tools, devices and platforms. Inauthentic audio is another growing trend and often hard to detect.

# 1. Introduction

## Background and motivation

AI is rapidly reshaping the information landscape across the UK and the number of claims appearing in social media and elsewhere online where the content is AI-generated or AI-altered is rising.[5] Personalised user experiences, combined with often low levels of media literacy and the ease of use of powerful generative AI tools leave existing defence mechanisms to information threats - such as libel laws and broadcasting regulations - outdated and largely ineffective. Increasing exposure to AI-generated and AI-adapted information makes it harder for people to trust what they see, read or hear,[6] with potential negative impacts on real-world behaviour and outcomes.[7]

---

[5] We use the term "AI-generated" to refer to content that is entirely or partially created - new - using artificial intelligence. This includes the creation of new text, still images and video. It also includes the creation of new audio, used over existing video. We use the term "AI-altered" to refer to existing content that is altered or manipulated to create a different meaning.

[6] Felix Simon, Rasmus Kleis Nielsen & Richard Fletcher, "Generative AI and news report 2025: How people think about AI's role in journalism and society," *Reuters Institute for the Study of Journalism*, 7 Oct 2025 https://reutersinstitute.politics.ox.ac.uk/generative-ai-and-news-report-2025-how-people-think-about-ais-role-journalism-and-society

[7] (i) Michael P Lynch, "Fake News and the Internet Shell Game," The New York Times, November 28, 2016, https://www.nytimes.com/2016/11/28/opinion/fake-news-and-the-internet-shell-game.html, for impact on public trust in information; (ii) For the effect of the perception of being disinformed has on trust in, and willingness to act on, accurate information, see: Michael Hameleers, Toni Van der Meer and Anna Brosius, "Feeling 'disinformed' reduces compliance with COVID-19 guidelines," *Harvard Misinformation Review*, May 31, 2020,

This report examines whether and how this public exposure to an increased volume of AI-generated content has caused or had potential to have caused or contributed to specific, real-world harms. The report draws on evidence identified in a dataset of 112 fact checks published by the leading UK fact checking charity, Full Fact, in 2025 and 2026, investigating AI-generated and AI-altered content circulating in the UK and around the world. The report identifies the topics, formats, and types of this AI-generated and AI-altered content, the actors who drive it, and the context in which it is produced. It provides evidence of the proven or potential consequences, or harms, which could be or could have been caused to individuals and society as a result of the content examined in the database. The report sets out who these consequences could have affected and, where possible, the potential severity and duration of these effects. The report details the model used to make the harm-risk assessment. This evidence is also available as a dataset that can be used by decision makers, civil society, and other researchers such as AISI grantees to create solutions that build resilience into the systems and infrastructure in which AI is deployed.

The aim of the report is to map and provide evidence of the proven or potential societal impact of AI-generated disinformation and misinformation, and the dynamics that propel these effects, and to provide evidence of the means and methodologies for risk monitoring in future, in line with AISI's research priorities.

## Research questions

Aligning with several of the research challenges set out in the AISI Priority Research Areas under Societal Resilience, including: (i) mapping the landscape; (ii) evidencing societal impact dynamics, and: (iii) investing in data, infrastructure and methodologies for risk monitoring, the report seeks to answer the following research questions:

- What range of types, formats and topics of AI-generated and AI-altered disinformation or misinformation were audiences in the UK exposed to in the UK in the study period?
- How does this differ, if at all, from the types, formats and topics of forms of disinformation and misinformation produced by other means?
- What is known of the size, and nature, of the audiences exposed to this disinformation and misinformation?

---

https://misinforeview.hks.harvard.edu/article/feeling-disinformed-lowers-compliance-with-covid-19-guidelines-evidence-from-the-us-uk-netherlands-and-germany/. (iii) for the effect of perceiving disinformation to be widespread on satisfaction with democracy, see: Andreas Jungherr & Adrian Rauchfleisch, "Negative downstream effects of alarmist disinformation discourse: Evidence from the United States" *Political Behavior*, 2024, 2123–2143. https://doi.org/10.1007/s11109-024-09911-3

- What evidence is there of the proven or potential consequences, or harms, for individuals or society of this AI-generated or AI-altered disinformation and misinformation?
- What would a content classification framework look like, based on the potential of different forms of AI-generated disinformation to cause or contribute to different forms, severity and duration of consequences, or harms, for individuals and society?

## Scope and limitations

The sample base for the project comprises fact checks and articles published by Full Fact that met the following criteria:

- Published between 1 January 2025 and 31 March 2026.
- The fact checks or articles mentioned the possibility that AI generation or manipulation played a role in the claim being checked. The content did not always establish this with complete certainty, and in some cases there was a possibility that the content was a human-made digital creation.

During the research period, there was no comprehensive tagging system in place to automatically compile a list of content meeting the above criteria. Researchers used a keyword site search for mentions of “AI” “AI-”, “Artificially” and “Artificial” to assemble a starting list and the fullfact.org domain was monitored continuously for new content up until the end of March 2026. A number of other pieces were added using reference to Full Fact’s internal metrics tracking which content was potentially AI-related but which did not explicitly mention AI in the final articles. The content was reviewed manually by researchers and false positives were de-selected during the entry process. A total of 112 published articles were eventually included in the project using this system.

Only a subset of the articles analysed established with certainty that AI was involved either wholly or partly in producing the false information. Certainty was not quantified in any of the checks considered, so a textual analysis was required to measure the degree of certainty established by the fact checks.

In 33 of the 112 cases, the fact check’s verdict was unequivocal that AI was involved in the content being checked. In a further 13 cases, the language claimed that AI was “almost certainly” involved, or a close equivalent of this phrasing. In the remaining cases, the language was weaker, finding either that AI’s role was “very likely”, “likely” or “possible”, or not directly assessing the likelihood of AI at all.

# 2. Literature Review

## Existing evidence on misinformation and disinformation harms

Academic debate over the effects of different forms of information on public opinion and behaviour dates back over a century to the decade after World War I.[8] And as research of misinformation and disinformation became "firmly entrenched" as an academic discipline over the past 10 years,[9] researchers have both found and debated evidence of the effects of different forms of false information in fields ranging from politics[10] to conflict[11] to public health[12] and the justice system.[13]

The fundamental step of determining what is and is not 'true', and whether and in what circumstances this information impacts audience understanding, opinion and behaviour has been debated by distinguished philosophers and writers, from Michel Foucault[14] and Bernard Williams[15] to Hannah Arendt[16] and Dan Sperber,[17] over many years. These debates often raise interesting questions about the definitions of different types of communication, different types of information, and the different ways that such information affects public understanding and attitudes.

---

[8] One of the first works in the field was a study of the effects on public opinion of propaganda emitted by the nations participating in World War I. Harold Lasswell "Propaganda techniques in the world war", *Knopf*, 1927. https://archive.org/details/dli.ernet.233727/page/13/mode/2up

[9] Chico Q Camargo and Felix M Simon, "Mis-disinformation studies are too big to fail: Six suggestions for the field's future". *Harvard Misinformation Review*, 20 September 2022, https://doi.org/10.37016/mr-2020-106.

[10] Richard Gunther, Paul Beck and Erik Nisbet, "Fake News May Have Contributed to Trump's 2016 Victory," *Ohio State University*, 2018, https://www.documentcloud.org/documents/4429952-Fake-News-May-Have-Contributed-to-Trump-s-2016.html.

[11] See, for example, this discussion of the British and Allied disinformation operations in World War II: Joshua Levine, "Operation Fortitude: The greatest hoax of the Second World War," Collins, 2012.

[12] See this study of the public health effects of false claims spread about a WHO polio vaccination campaign in Nigeria in 2002: Ayodele Jegede, "What Led to the Nigerian Boycott of the Polio Vaccination Campaign?" *PLOS Medicine*, 2007, https://doi.org/10.1371/journal.pmed.0040073.

[13] Academics have studied the effect of misinformation in the criminal justice system for many years. See: Steven Frenda, Rebecca and Elizabeth F. Loftus, "Current Issues and Advances in Misinformation Research," *Current Directions in Psychological Science,* 2021. https://doi.org/10.1177/0963721410396620.

[14] Michel Foucault, "The Order of Things", *Routledge Classics*, 2002. First published 1966.

[15] Bernard Williams, "Truth and truthfulness. An essay in genealogy," *Princeton University Press*, 2002.

[16] Hannah Arendt, "Lying in Politics", in "Crises of the Republic*", Harvest*, 1972.

[17] Dan Sperber, "Apparently irrational beliefs" in *Rationality and Relativism*, ed. Martin Hollis and Steven Lukes, MIT Press, 1982, pp 171-172, https://www.dan.sperber.fr/wp-content/uploads/1982_apparently-irrational-beliefs.pdf.

As researchers Emily Vraga and Leticia Bode noted in 2020, defining misinformation and disinformation in a "consistent and coherent way" has long been a challenge.[18] However, in more recent years, greater academic consensus has grown around these key terms. The term 'misinformation' is broadly accepted by academics today as referring to information - shown by the best evidence publicly available at the time to be false or misleading - which is not known to be false by those who create and spread it.[19] 'Disinformation' is used, by contrast, for referring to false or misleading information that is known to be so by those who create and spread it; most often for political or social effect. In addition, academics identify other forms of false information created with intent to mislead - but for either financial reward or other forms of gratification; not for political or social effect. This includes both "scams" or "hoaxes" and "clickbait."

While a broad consensus has grown on the meanings of these terms, if not the way they are sometimes used,[20] academics have for decades debated the nature of communication; some seeing it primarily in terms of imparting basic factual understanding and others as primarily a means for expressing broader attitudes, opinions and beliefs.[21] At the same time, cognitive scientists and psychologists have researched the factors that shape how different forms of communication affect both understandings and attitudes. Factors assessed include the way existing views shape the effect of information on belief,[22] how easy the information is to process[23] and the effects of claim repetition and false narratives on belief.[24] Political and social scientists have explored these and other factors shaping the persistence of false factual understandings and beliefs.[25]

---

[18] Emily Vraga and Leticia Bode, "Defining Misinformation and Understanding its Bounded Nature: Using Expertise and Evidence for Describing Misinformation" *Political Communication,* Vol 37(1) 136-144, 2020, https://doi.org/10.1080/10584609.2020.1716500.

[19] It should be noted that while different terms - misinformation and disinformation - exist in English, distinguishing between false and misleading information that is or is not known by those who spread it to be false, this distinction is not recognised in some other languages including French and Spanish.

[20] The broad academic consensus around the meaning of the term does not, of course, always extend to agreement on how the term is applied, with many of those accused of spreading false information rejecting the claim.

[21] See: (i) Dan Sperber, "Apparently irrational beliefs" in *Rationality and Relativism*, ed. Martin Hollis and Steven Lukes, MIT Press, 1982, pp 171-172, https://www.dan.sperber.fr/wp-content/uploads/1982_apparently-irrational-beliefs.pdf. (ii) James Carey, "Communication as Culture: Essays on media and society," *Unwyn Hyman,* 1989.

[22] Mathias Osmundsen, Alexander Bor, Vahlstrup, Peter Bjerregaard, Anja Bechmann and Michael Bang Petersen, "Partisan Polarization Is the Primary Psychological Motivation behind Political Fake News Sharing on Twitter," *American Political Science Review*, 115.3 (2021), 999-1015, https://doi.org/10.1017/s0003055421000290.

[23] See: (i) Thomas E. Powell, Hajo G. Boomgaarden, Knut de Swert and Claes H. de Vreese, "A clearer picture: The contribution of visuals and text to framing effects," *Journal of Communication*, 65.6, (2015), 997–1017, https://doi.org/10.1111/jcom.12184. (ii) Paul Messaris and Linus Abraham, "The role of images in framing news stories", in *Framing public life. Perspectives on Media and Our Understanding of the Social World,* edited by Stephen Reese, Oscar Gandy and August Grant, (New Jersey & London: Lawrence Erlbaum Associates Publishers, 2001), 215–226.

If some broad level of agreement exists on factors shaping the effects of information on understanding and attitude, the most significant debate remaining within the field relates to the effects of this information on behaviour.

Broadly, the argument that has emerged over the past decade follows the same pattern as the debate about "media effects" that developed over the past century after, in the wake of World War I, political and social scientists in the United States identified the mass media of the time - film, radio and print press - as having "powerful effects" in shaping public opinion and behaviour.[26] This view held until, during World War II two decades later, a new generation of researchers emerged whose studies showed the media as having only "limited effects" on the public's opinions and behaviour. A series of these new studies published in the 1940s onward identified other, social, factors as having greater influence than media on attitudes and behaviour.[27] This view also held for several decades until, from around the 1970s onward, while many continued to reject the notion of media having "powerful effects", other researchers recognised that such effects do or may occur with certain audiences and in certain circumstances and that understanding how this happens required refining academics' theoretical explanations.[28]

In a similar way today, three broad schools of thought emerged over the past decade, after the political, social upheavals of 2016 that caused a surge in interest in the field. First to emerge was the school that saw misinformation and disinformation having significant power to reshape public opinion and behaviour - seen as causing or contributing to the election of Donald Trump in the United States and Rodrigo Duterte in

[24] See: (i) Floyd Allport and Milton Lepkin, "Wartime Rumors of Waste and Special Privilege: Why Some People Believe Them," *The Journal of Abnormal and Social Psychology*, 40.1, (1945), https://psycnet.apa.org/record/1945-01987-001. (ii) Lisa Fazio, Raunak Pillai and Deep Patel, "The effects of repetition on belief in naturalistic settings," *Journal of Experimental Psychology*, 151.10, (2022): 2604-2613, https://pubmed.ncbi.nlm.nih.gov/35286116/. (iii) Doris Lacassagne, Jeremy Bena, Olivier Corneille, "Is Earth a perfect square? Repetition increases the perceived truth of highly implausible statements," *Cognition*, Vol 223, (2022), https://doi.org/10.1016/j.cognition.2022.105052.

[25] See for example: Brendan Nyhan and Jason Reifler, "When corrections fail: The persistence of political misperceptions", Political Behavior, 32(2), 303–330 (2010). https://doi.org:10.1007/s11109-010-9112-2.

[26] See, for example: (i) Harold Lasswell, "Propaganda technique in the world war," *Knopf*, 1927. (ii) Edward Bernays, "Propaganda," *Horace Liveright*, 1928.

[27] See, for example, (i) Paul Lazarsfeld, Bernard Berelson and Hazel Gaudet, "*The people's choice,*" *Columbia University Press*, 1948; (ii) Joseph Klapper, *"The effects of mass communication,"* Free Press, 1960.

[28] Elizabeth M, Perse, "Media effects and society," *Lawrence Erlbaum Associates Publishers*, 2001.

The Philippines[29]. This view of events was swiftly followed by a second framing that identified misinformation and disinformation as causing little more than a false factual understanding, with "limited or zero" effects on public attitudes and actions.[30] Following from this view was a school that identified alarmist discourse about the threat misinformation poses as both negatively affecting confidence in the state of democracy and increasing support for restrictions on freedom of speech.[31] Meanwhile, the third framing that emerged identified particular forms of false information as having potential to cause effects in particular circumstances, and not in others, in a broad range of fields. The different fields identified in this research range from politics[32] to social attitudes[33] to vigilante violence,[34] to individual well-being,[35] to unrest and conflicts,[36] to public health[37]

---

[29] See: (i) Max Read, "Donald Trump Won Because of Facebook," *New York Magazine*, 9 November 2016, https://nymag.com/intelligencer/2016/11/donald-trump-won-because-of-facebook.html, from popular media, and studies that supported this view, (ii) Richard Gunther, Paul Beck and Erik Nisbet, "Fake News May Have Contributed to Trump's 2016 Victory," *Ohio State University*, March 2018, https://www.documentcloud.org/documents/4429952-Fake-News-May-Have-Contributed-to-Trump-s-2016.html; (iii) Martin Moore, "Democracy Hacked: How Technology is Destabilising Global Politics*" One World Publications*, 2018; (iv) Ullrich Ecker, Jon Roozenbeek, Sander van der Linden, Li Qian Tay, John Cook, Naomi Oreskes and Stephen Lewandowsky, "Misinformation poses a bigger threat to democracy than you might think", *Nature*, June 5, 2024, https://www.nature.com/articles/d41586-024-01587-3.

[30] See (i) Joshua Kalla and David Broockman, "The minimal persuasive effects of campaign contact in general elections: Evidence from 49 field experiments," *American Political Science Review*. 28 September 2017, https://doi.org/10.1017/S0003055417000363. (ii) Andrew Guess, Dominique Lockett, Benjamin Lyons, Jacob Montgomery, Brendan Nyhan and Jason Reifler, "'Fake news' may have limited effects beyond increasing beliefs in false claims," *HKS) Misinformation Review*, 2020. https://misinforeview.hks.harvard.edu/article/fake-news-limited-effects-on-political-participation/. (iii) Ceren Budak, Brendan Nyhan, David M Rothschild, Emily Thorson and Duncan J Watts, "Misunderstanding the harms of online misinformation," *Nature*, 630 45-53 (2024), https://doi.org/10.1038/s41586-024-07417-w.

[31] Andreas Jungherr & Adrian Rauchfleisch, "Negative downstream effects of alarmist disinformation discourse: Evidence from the United States" *Political Behavior*, 2024, 2123–2143. https://doi.org/10.1007/s11109-024-09911-3

[32] Stephanie Burchard, "Electoral violence in sub-Saharan Africa. Causes and consequences," *Lynne Rienner*, 2015, p18.

[33] Florian Foos and Daniel Bischof, "Tabloid Media Campaigns and Public Opinion: Quasi-Experimental Evidence on Euroscepticism in England," *American Political Science Review*, 2021, 19-37, https://doi.org/10.1017/S000305542100085X.

[34] Shakuntala Banaji, Ram Bhat, Anushi Agarwal, Nihal Passsanha and Mukti Pravin, "WhatsApp vigilantes: An Exploration of citizen reception and circulation of WhatsApp misinformation linked to mob violence in India," London School of Economics, 2019, https://www.lse.ac.uk/media-and-communications/assets/documents/research/projects/WhatsApp-Misinformation-Report.pdf.

[35] Samantha Brooks and Neil Greenberg, "Psychological impact of being wrongfully accused of criminal offences: A systematic literature review," *Medicine, Science & the Law*, 61.1, (2020), https://doi.org/10.1177/0025802420949069.

[36] See: (i) Joshua Levine, "Operation Fortitude: The greatest hoax of the Second World War,*"* Collins, 2012. (ii) Simone Monasebian, **"**The Pre-Genocide Case Against Radio-Télévision Libre des Milles Collines in **"***The media and the Rwanda Genocide",* edited by Allan Thompson, 308-330, *Pluto Press*, 2007.

[37] Nathan Thielman, Jan Ostermann, Kathryn Whetten, Rachel Itemba, Dafrosa Itemba, Venance Maro, Brian Pence and Elizabeth Reddy, "Reduced Adherence to Antiretroviral Therapy among

and health policy,[38] to the functioning of the justice system,[39] the fortunes of particular businesses, markets and the economy[40] responses to climate change,[41] and more - depending on the nature of the information, the audience and context. And this whole body of research - from powerful, to limited to particular effects - helps inform the findings in this study.

## AI's role in the information landscape

Fact checks which confirm or suggest AI's involvement have become more common in Full Fact's output and account for an increasing proportion of misinformation monitored. Full Fact's first check involving AI-generated content took place in early 2023, similar to other fact checkers.[42] In the study period alone, confirmed or suspected AI claims accounted for 24% of all fact checks published. Had the same study been conducted a year previously, it would only have made up 8% of content.

Experience shows that both improvements in generative AI tools and their wider availability have lowered the barrier for creating misleading content at scale. It is also posing new challenges for fact checkers and journalists in successfully detecting and proving that AI has played a role. The ease of rebuttal varies case-by-case but, in line with the growing sophistication of AI tools, many content types are becoming more difficult to detect. Glitches such as inaccurate limbs or body parts on people—which were frequent three years ago—are still the most common means of rebuttal but identified less often than before.

Deepfakes—when a person's likeness is copied in order to produce inauthentic behaviour or speech—have become one of the most common and recognisable forms of

---

HIV-infected Tanzanians Seeking Cure from the Loliondo Healer," *Journal of Acquired Immune Deficiency Syndrome*, 2014, https://doi.org/10.1097/01.qai.0000437619.23031.83.

[38] See (i) Ayodele Jegede, "What Led to the Nigerian Boycott of the Polio Vaccination Campaign?" *PLOS Medicine*, 2007, https://doi.org/10.1371/journal.pmed.0040073. (ii) Pride Chigwedere, George Seage, Sofia Gruskin, Tun-Hou Lee and M Essex, "Estimating the lost benefits of antiretroviral drug use in South Africa," *Journal of Acquired Immune Deficiency Syndrome*, 2008, 410-415, https://doi.org/10.1097/qai.0b013e31818a6cd5.

[39] See (i) Elizabeth F. Loftus, "Planting misinformation in the human mind. A 30-year investigation of the malleability of memory," *Learning and memory*, 2005, 361-366, https://doi.org/10.1101/lm.94705; (ii) Steven Frenda, Rebecca and Elizabeth F. Loftus, "Current Issues and Advances in Misinformation Research," *Current Directions in Psychological Science,* 2021. https://doi.org/10.1177/0963721410396620.

[40] Martin Baily, Robert Litan and Matthew Johnson, "The Origins of the Financial Crisis," *Brookings Institute*, November 2008, https://www.brookings.edu/wp-content/uploads/2016/06/11_origins_crisis_baily_litan.pdf.

[41] Michael Tesler, "Elite Domination of Public Doubts About Climate Change (Not Evolution)," *Political Communication*, 2017,306-326. https://doi.org/10.1080/10584609.2017.1380092.

[42] Grace Abels, "Can ChatGPT fact-check? We tested," *Poynter*, 31 May 2023, https://www.poynter.org/fact-checking/2023/chatgpt-ai-replace-fact-checking/.

AI-generated disinformation. In many cases the rebuttal process is straightforward, as the authentic video used as a template can be found and compared to the fake footage. However content that is exclusively audio has consistently proven more challenging for fact checkers to identify with certainty. This is a combination of experts being unable to reach definitive conclusions from their own tools and experience, and the possibility that skilled impersonators, rather than AI-generated voices, are used instead.

New forms of detection are challenging this trend. SynthID[43]— invisible watermarks which indicate that an image or video still have been created or altered using specific AI tools—has become increasingly relied on in AI checks. Google introduced SynthID in 2023 and has made advanced detection tools available to some fact checkers and researchers.[44] OpenAI began a similar programme in 2026 which is already being detected through fact checks.[45],[46]

## Current detection and classification approaches

In response to the rapid rise in the volume and impact of AI-generated content there are many approaches to its detection. When early photo-realistic AI-generated content was first shared online, closer inspection usually revealed telltale signs it had been faked. For example, the famous 2023 image of Pope Francis in a puffer jacket is superficially convincing, but zooming in on his hands reveals glitches.[47] This viral image marked a tipping point: most earlier images were typically fantastical and not attempting to be realistic, but since then, they have become harder to identify by simply examining the content of the image.

Since 2020 a series of initiatives aimed at making it easier to identify AI-generated content have been developed.

The Coalition for Content Provenance and Authenticity (C2PA) defines a standard and secure way to attach "Content Credentials" to digital media, allowing content creators to share who create the media, how it has been edited and shared. For example, if a photograph is taken with a digital camera and then genAI is used to enhance the image by cleaning the background, the content credentials can capture where and when the photo was taken, and exactly how it has been edited. This information can then be verified when the image is shared so its provenance can be verified, even if some aspect of the image is AI manipulated.

[43] Google Deepmind, "A tool to watermark and identify content generated through AI," Undated, https://deepmind.google/models/synthid/.
[44] Laurie Richardson & Pushmeet Kojhli, "Making it easier to understand how content was created and edited," Google Blog, 19 May 2026, https://blog.google/innovation-and-ai/products/identifying-ai-generated-media-online/.
[45] OpenAI, "Advancing content provenance for a safer, more transparent AI ecosystem," 19 May 2026, https://openai.com/index/advancing-content-provenance/.
[46] Sian Bayley, "AI images from Beaver Scouts mosque visit circulate online," *Full Fact*, 10 June 2026, https://fullfact.org/culture-and-society/ai-image-beaver-scouts-mosque-visit/.
[47] "Twitter users fooled by AI images of Pope in street fashion", CEDMO, 2023 https://cedmohub.eu/twitter-users-fooled-by-ai-images-of-pope-in-street-fashion/

As well as following the C2PA standard, Google has developed its own "SynthID" system to add an invisible watermark to media generated from its AI tools as mentioned earlier. Rather than being a piece of metadata attached to the media, SynthID is embedded in the media itself. Google provides controlled access to an API that can verify if a piece of media has such a watermark. Recently, Google announced partnerships with Nvidia and OpenAI to introduce genAI watermarks across multiple tools and platforms.[48] For several years, the EU's Horizon grants funded "vera.ai", a research and development project that analysed disinformation and developed AI tools to verify media. While the project has now finished, its tools remain available[49], including those for synthetic image and audio detection.

Of course, while such initiatives can assist fact checkers and others to identify AI-generated information, a weakness is that they are neither widely understood nor used by members of the public.

# 3. Methodology

## Claim selection criteria (checkworthiness framework)

Fact checkers do not always know at the outset whether a given piece of content is true, false or misleading, whether it is mis- or disinformation, or whether AI has played a role in its creation. Full Fact and other fact checking outlets use a process of monitoring assisted by detection tools for identifying checkable factual claims from the news media, politics and social media.[50] The decision to check a particular factual claim takes into account a number of factors[51] listed below:

---

[48] Ryan Whitwam, "Google's SynthID AI watermarking tech is being adopted by OpenAI, Nvidia, and more," Ars Technica, 19 May 2026, https://arstechnica.com/google/2026/05/googles-synthid-ai-watermarking-tech-is-being-adopted-by-openai-nvidia-and-more/
[49] Vera.Ai, "Vera.ai has come to an end – its results remain accessible. More right here!", Undated, https://www.veraai.eu/posts/veraai-final-project-results.
[50] "Full Fact AI", *Full Fact*, Undated, https://fullfact.org/ai/
[51] Peter Cunliffe-Jones and Andrew Dudfield, "Spotting the harmful needles in the haystack: How fact-checkers select what to check," *Poynter*, 5 August 2025, https://www.poynter.org/ifcn/2025/spotting-the-harmful-needles-in-the-haystack-how-fact-checkers-select-what-to-check/.

- Claims on a matter of contemporary debate, related to a topic concerning the public interest, typically made by major politicians and media outlets.
- Claims on topics that may have the potential to negatively affect peoples' lives.
- Claims that are either widely shared online or have potentially high reach via the print and broadcast media, and political candidates, and where there may be a public interest in quickly rebutting or challenging the content.
- Claims which have educational value as case studies for certain forms of mis- or disinformation or to exemplify tactics used by those who spread it.
- Claims on Facebook and Instagram flagged by users as potential misinformation and eligible for inclusion in the Third Party Fact Checking programme.

Fact checkers do not publish the results of every investigation. Some projects that are inconclusive or fail to uncover misinformation are not pursued beyond initial research. This creates a bias in published content towards claims that have been found to be false or misleading, and towards more definitive cases. The upshot of this approach and checkworthiness framework is that fact checkers do not check a randomised sample of all factual claims made in public debate, but rather focus on claims considered worth checking. This is an important caveat for analytical purposes, as it limits the value of generalising the findings from a set of fact checks to make more general pronouncements about information trends.

## AI detection methods

Each claim in the study was analysed to establish the role that AI played in the creation of the content and the models or platforms involved in producing it.

Glitches were the most common type of evidence establishing the role of AI; identified in 68 of the 112 entries in the study. Glitches are defined as imperfections in the rendering of visual or audio items. For example, in an image purporting to depict Iranian soldiers escorting US servicemen away from the site of a crashed stealth bomber, one of the soldiers is shown with three hands.[52] In another example, an audio clip claiming to present a newsreader reporting a story about Prince William and London Mayor Sadiq Khan glitches included unnatural pauses and an unusual cadence in the dialogue.[53]

The second largest category of evidence was incongruities, present in 63 cases. Incongruities are elements which don't cohere with a verified reality. For example, one clue that images of Venezuelan President Maduro in US custody were AI-generated was that he was shown wearing different clothing than in verified images that showed the reality.[54] Identifying incongruities is a key element of debunking AI-altered and deepfake content, by referencing the real videos they are mimicking.[55]

---

[52] Hannah Smith, "Image showing US B-2 supposedly 'shot down' by Iran is fake", *Full Fact*, 11 March 2026, https://fullfact.org/conflict/fake-B-2-shot-down-image/
[53] Charlotte Green, "Prince William has not challenged Sadiq Khan over London's priorities", *Full Fact*, 26 February 2026, https://fullfact.org/politics/prince-william-sadiq-khan-london-veterans-housing-false/

The third most common category was "negative evidence", or an absence of credible corroborating evidence, to substantiate a claim. This applied in 60 cases. An example of this is an image purporting to show a satellite which had been brought down by an Iranian missile. As the fact check noted: "there have been no reports of any satellite being targeted by Iran in the way the posts claim. Any such attack on another country's satellite is likely to have created significant space debris that is detectable by private companies and governments."[56]

The next category of evidence was recognition of evidence bearing resemblance to previous debunks, in terms of subject matter or a trope of disinformation design, which appeared in 47 cases. For example, many of the claims analysed were on the subject of personal freedoms being infringed, and the content used similar formatting, backgrounds and voiceover styles, suggesting a likelihood of either a common creator or common creation platform.[57],[58],[59]

Watermarks were present in 36 of the entries analysed, and these provided the strongest evidence of the artificial origins of the content. Half of these cases (18) were Google Gemini watermarks, of which 13 were only detectable using SynthID. The other five contained a visible watermark on the content. Grok watermarks were shown in four cases, Sora in three cases, and one each of Runway, InShot and Unreal. The remaining eight cases contained a written disclaimer or user's name which did not specify the AI tool used.

The next category was subject testimony, which provided evidence in 32 entries. This involved testimony or confirmation from either the subject of the claim, or an interested party. A fairly common use of this evidence type was in apparent scam content in which the creator sought to imitate a retailer. In these cases, the retailer's own denial was taken as strong evidence that the content was not real.[60]

---

[54] Sian Bayley, "Viral image of Nicolás Maduro being led away from plane isn't real," 7 January 2026, https://fullfact.org/world/picture-maduro-seized-us-military-ai/

[55] Sian Bayley, "Another AI-generated video of Donald Trump criticising Keir Starmer circulates online," *Full Fact*, 30 January 2026, https://fullfact.org/us/donald-trump-criticising-keir-starmer-ai-2/

[56] Charlotte Green, "Fake crash picture shared with false claims it shows satellite brought down by Iranian missile," *Full Fact*, 20 March 2026, https://fullfact.org/conflict/satellite-israel-iran-fake/.

[57] Leo Benedictus, "You won't be charged £300 for using too much electricity after 9pm," *Full Fact*, 28 November 2026, https://fullfact.org/environment/fake-300-charge-after-9pm/.

[58] Joshua Salisbury, "No TV licence exemption for the over 60s," *Full Fact*, 28 November 2026, https://fullfact.org/economy/tv-licence-exemption-over-60-false/

[59] Hannah Smith, "Heathrow and Gatwick aren't introducing £50 airport entry fees," *Full Fact*, 4 December 2025, https://fullfact.org/economy/heathrow-gatwick-entry-fees/.

The testimony and analysis of experts was used to confirm or point to the use of AI in 19 of the cases reviewed. This included testimony from Professor Hany Farid at the University of California, Berkeley and Dr Siwei Lyu at the University at Buffalo.

The final two significant forms of evidence were analysis of the track record of the accounts sharing the content (18 entries) and admission by the account that it used AI (11 cases). Some accounts involved declared that they were satirical or contained obvious joke descriptions.[61]

## Harm risk assessment model

The harm risk assessment model used to establish the findings here[62] was developed in a four-year research project, explained in a peer-reviewed study published by University of Westminster Press, in 2025.[63]

The project identified a set of factors shaping the potential of specific examples of false or misleading information to cause or contribute to negative effects, or 'harms', for individuals and society by examining: (i) A dataset of 250 specific examples of misinformation and disinformation - to understand the factors that shaped which did and which did not have potential for significant consequences; (ii) Published evidence of the nature and drivers of effects of a range of different forms of false information in fields ranging from politics and business to public health and the justice system.

To be found to have a substantive potential to cause or contribute to particular effects, evidence must show that entries in the dataset both (i) meet agreed criteria on factors set out in the model and (ii) earlier research must show similar claims to have caused, or contributed to similar effects in similar circumstances in the past.

Used since 2024 in a series of trials and studies in Europe, Africa, and the Middle East, the harm risk assessment model recognizes the potential of false information to cause or contribute to both specific and broad substantive effects, or consequences. The model identifies substantive consequences as changes causing objectively verifiable substantive injuries to the interests of an individual, group, or wider society - not causing a false understanding nor causing a fleeting negative emotion such as fear or anger.

---

[60] Sian Bayley, "No, Cadbury isn't selling an 'Eid egg'," *Full Fact*, 26 February 2026, https://fullfact.org/culture-and-society/cadbury-not-selling-eid-egg/

[61] Twitter account: https://x.com/BotFinderUK

[62] Peter Cunliffe-Jones, "Fact checking what matters: How a harms-based model for selecting claims works", *Harvard Kennedy School (HKS) Misinformation Review,* July 2026. https://doi.org/10.37016/mr-2020-197.

[63] Peter Cunliffe-Jones, "Fake news: What's the harm?" 2025, *University of Westminster Press*. https://doi.org/10.16997/mpub.14614695

As set out below in Figure 1, the model determines risk of harm based on answers to three broad questions:

I. Whether the claim, if believed, causes or contributes to a substantively false factual understanding, or whether the understanding caused is only narrowly inaccurate,[64]

II. Whether – if the claim is found to cause a substantively false understanding – evidence shows it is believed by sufficient people to cause one or more of the specific consequences identified in the model as having potential to be caused by that type of claim,

III. Whether – if the claim is found to cause a substantively false understanding and it found to be believed by sufficient people – evidence shows that those who believe the claim have, or may have, the capacity and motivation to act on the false understanding it causes either now, or in future if this and similar claims are repeated over time.

Depending on the answers, the model finds there is or is not a substantive potential for consequences and helps determine the type of field of effect and, in some cases, its scale, severity and duration.

To consider these three questions, the fact-checkers at Full Fact applied the model by examining evidence related to 14 points from the topic or topics of the claim to: the degree of falsity in the claim; the size and nature of the audience; audience perceptions of claim credibility; whether the claim or similar claims had been repeated over time; whether the claim fit into a broader false narrative or conspiracy theory, and whether those who believed the claim had the capacity and may have been motivated to act on the factual understanding or attitude caused.

[64] See the Glossary at the end of this report for the definitions of all key terms used here .

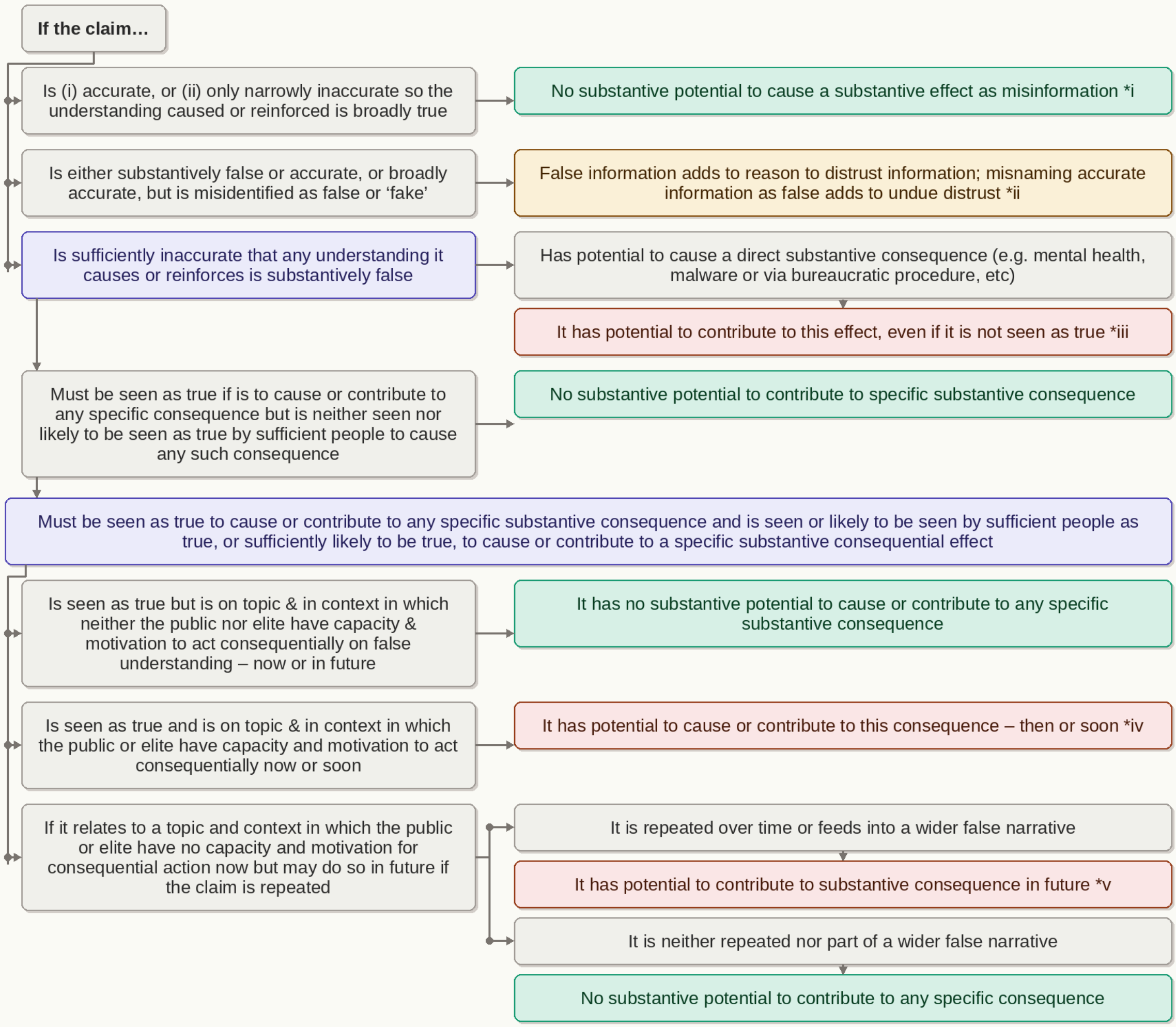

*Figure 1. How three broad factors — (i) the degree of falsity, (ii) perceived accuracy & (iii) audience capacity & motivation to act — shape potential for substantive consequences.*

If the evidence recorded showed (i) that the claim was substantively false, (ii) that it was believed by sufficient people to cause or contribute to a particular consequence, and (iii) that sufficient of those who believed the claim had both the capacity and potential motivation to act on the understanding caused, now or in future, the model indicates the type or types of consequence that such claims are known to have caused or contributed to as a potential effect. It then asks the user to consider who the consequences might affect - from a specific individual, to non-specific individuals in general, to a specific group or community, to a large part or the whole of society, and, in some cases, the potential severity and duration of these effects. In addition the model identified whether the claim has potential for direct effect – through action taken in the near-term – or by contribution to the cumulative effect of this and similar false claims on an audience's attitude or understanding over time.

An independent study of the model by doctoral researchers at the University of Wisconsin-Madison, made public in 2025, recognised that the judgements required in predicting potential effects of false information are complex and this could lead to inconsistencies in findings.[65] To ensure consistency in how the model was applied for this study, we used a gold standard review of the findings produced by the team, to verify that the evidence showed the criteria had been met. By this process, we determined the potential consequences of the misinformation and disinformation in the dataset.

---

65 CAMRI. "Trial finds predictive model helps fact checkers identify false claims with potential to cause harm," *Communications & Media Research Institute*, 2 April, 2025,. https://camri.ac.uk/blog/2025/04/02/trial-finds-predictive-model-helps-factcheckers-identify-false-claims-with-potential-to-cause-harm/.

# 4. Findings: The Dataset

## Overview

The dataset comprises 112 pieces of AI-generated or AI-altered content that circulated online between 1 January 2025 and 31 March 2026, and were seen tens of millions of times between them. The content in the dataset covered 19 broad topics - from business and the economy to health to politics to war and conflict. Most claims examined in the dataset related to 2-4 topics, for example concerning both politics and the economy or both the environment and transport.

The false information spread in a variety of formats - primarily video, a combination of video-and-audio, or still images. Two entries featured text created by an AI chatbot. Given both the growing use of AI chatbots by millions of people every day as a source of information and the number of errors produced by these tools,[66] text produced by AI chatbots could be an important area of further research.

Studies of misinformation have in recent years identified multiple mechanisms by which online and offline content may mislead audiences.[67] These range from satire that is mistakenly seen as true, to content that is fabricated or doctored to create a false understanding. In the dataset for this study, content fell primarily into two categories of information manipulation: (i) imposter content, i.e. information that impersonates an individual or group to create a false understanding of their views, character or actions; and (ii) fabricated content, i.e. images of a scene or event fabricated in a way that creates a false understanding of the situation or event. Studies of non-AI-generated misinformation typically show a broader range of formats.

All the entries in our study were in some way false or misleading. Applying the criteria set out in the model, our analysis identified 18 (or 16.1%) of the 112 entries as creating an understanding that was only narrowly inaccurate and 94 (or 83.9%) as creating a

[66] See: (i) Jamie Hancock & Azzurra Moores, "Electoral Hallucinations: Safeguarding UK elections in the world of LLMs and AI chatbots," *Demos*, 20 May 2026, https://demos.co.uk/research/electoral-hallucinations-safeguarding-uk-elections-in-the-world-of-llms-and-ai-chatbots/ (ii) And this report finding 1 in 10 men in the UK use AI chatbots such as ChatGPT for health information. "Men's Health: How to improve health outcomes, knowledge and behaviours" https://nds.healthwatch.co.uk/reports-library/mens-health-how-improve-health-outcomes-knowledge-and-behaviours

[67] Claire Wardle, "Understanding information disorder," *First Draft*, 22 September 2020, https://firstdraftnews.org/long-form-article/understanding-information-disorder/#

substantively false or misleading understanding.[68]

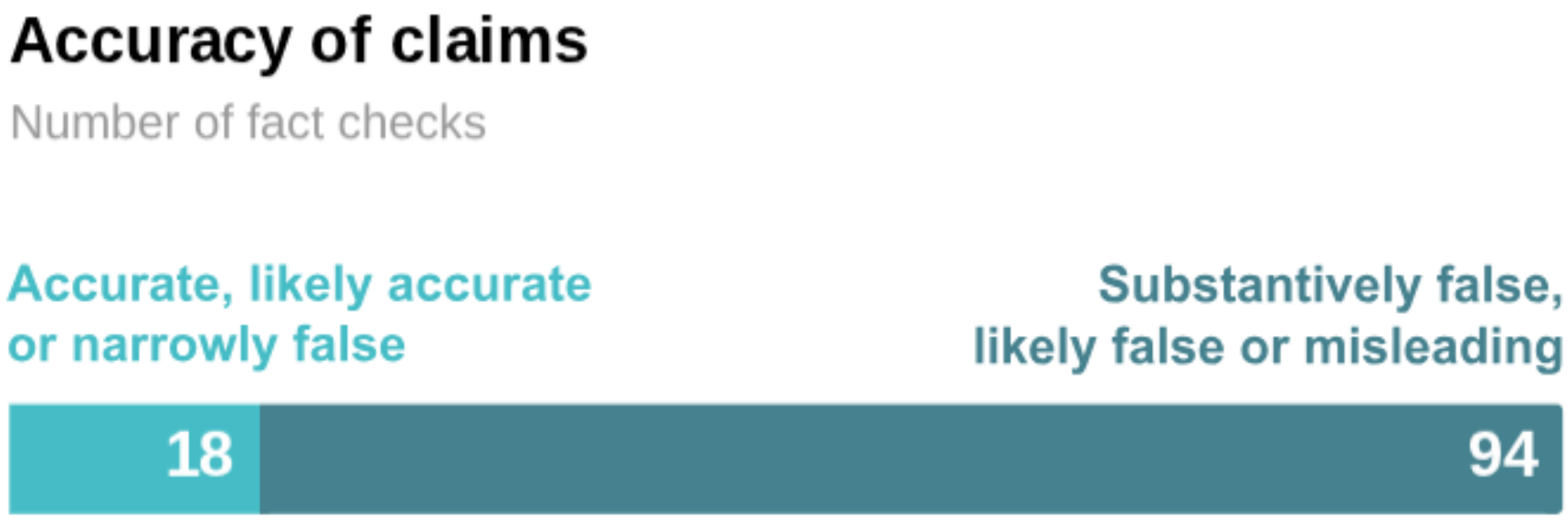

In terms of topics, the most common topics of content appearing in the dataset related to what could be broadly described as “politics”; a theme appearing in 71 (63.4%) of the 112 dataset entries. Much of this appeared in the form of AI-generated video or audio purporting to show prominent politicians or establishment figures making, such as the then prime minister or the mayor of London, making announcements of supposed government policies which they had not made.

[68] The model identifies claims as substantively false or misleading where evidence shows that (i) elements of the claim that are key to the understanding it creates are inaccurate (ii)elements of the claim that are true are either (a) distorted or out of context (b) unimportant. It identifies claims as only narrowly inaccurate where the elements that are inaccurate are either (i) immaterial to the understanding created, or (ii) so marginally inaccurate the broad understanding created would be essentially the same as created by an entirely accurate claim.

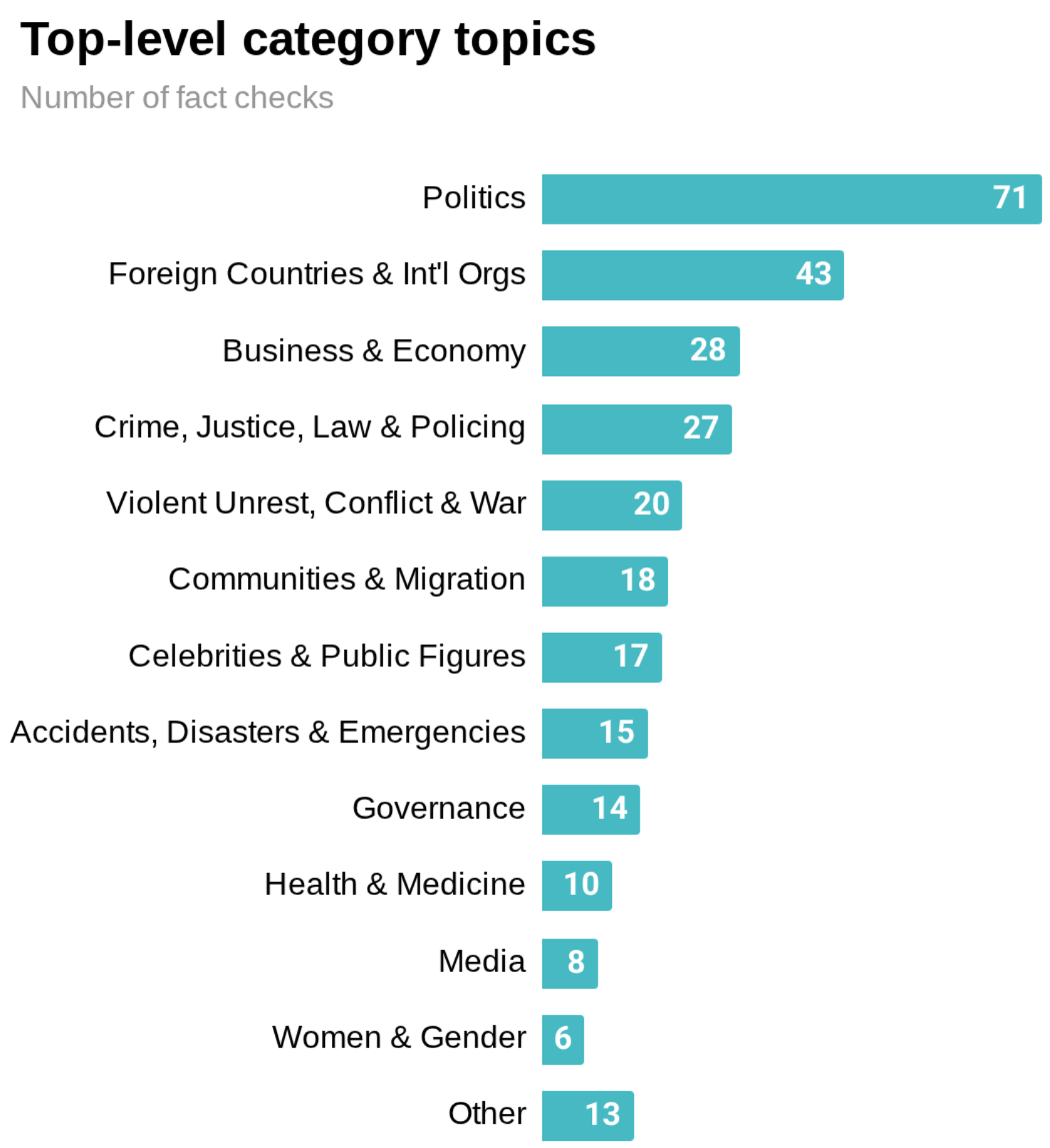

Other common themes included AI-generated disinformation about: (i) general events in foreign countries; (ii) events specific to the wars in the Middle East and Ukraine; (iii) changes to the individual economy and cost of living; (iv) the activities of the police or criminal justice system; (v) the actions of particular ethnic or religious communities; (vi) well-known conspiracy theories and (vii) claims related to health and medicine. With the exception of conspiracy theory content, a strong correlation could be seen between themes in the dataset and events, people or situations appearing in mainstream news.

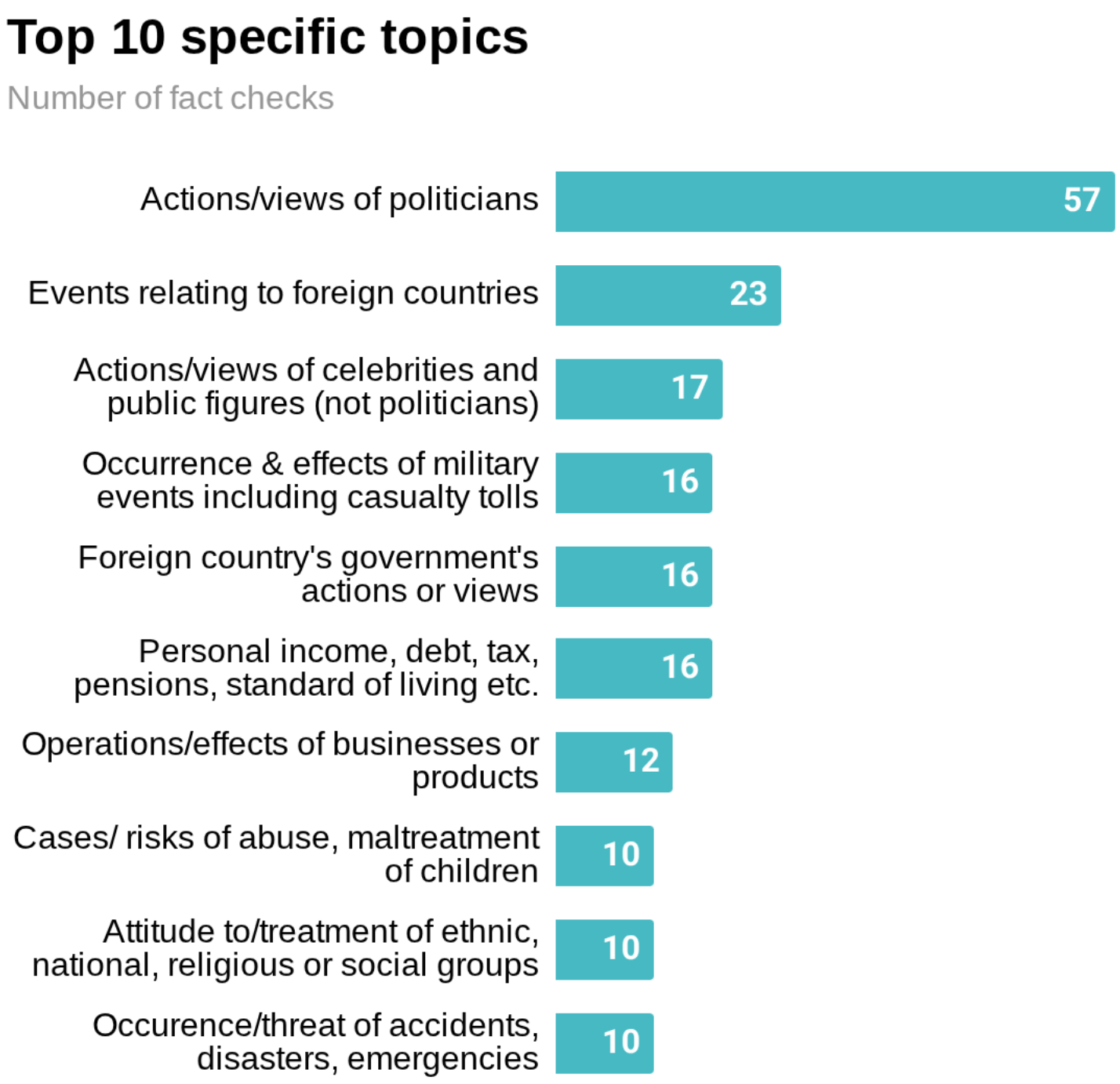

## Detail: key topics, themes & narratives in AI-generated dataset

As discussed above, the dataset is not a randomised sample of all AI-generated content online.[69] The topics, themes and narratives identified reflect the criteria set out above by Full Fact for "checkworthiness", whether that relates to events in the news or information such as scams and health-related misinformation with clear potential for consequences for individuals or groups.

In this way, the study identified content in the dataset as relating to one or more of 19 broad topics that are identified within the harm risk assessment model.

**Table 1: topics covered**

[69] Aisha Down, "More than 20% of videos shown to new YouTube users are 'AI slop', study finds," *The Guardian*, 27 December 2025, https://www.theguardian.com/technology/2025/dec/27/more-than-20-of-videos-shown-to-new-youtube-users-are-ai-slop-study-finds

| Broad topic | Fact checks |
|---|---|
| Politics | 71 |
| Foreign Countries & Int'l Orgs | 43 |
| Business & Economy | 28 |
| Crime, Justice, Law & Policing | 27 |
| Violent Unrest, Conflict & War | 20 |
| Communities & Migration | 18 |
| Celebrities & Public Figures | 17 |
| Accidents, Disasters & Emergencies | 15 |
| Governance – incl. costs, powers of state, abuse of powers + corruption, conspiracies | 14 |
| Health & Medicine | 10 |
| Media | 8 |
| Women & Gender | 6 |
| AI | 3 |
| Education | 2 |
| Environment & Climate | 2 |
| Actions of specific private individuals | 1 |
| Scams | 1 |
| Development | 1 |
| Science | 1 |

Three topics identified in other studies using the same model that did not appear in this dataset were claims related to (i) mental health (e.g. false statements about prevalence or causes of mental illness, as seen in other studies); (ii) faked content purporting to show the sexual behaviour of private and public individuals, or false claims related to sexuality and sexual health, and (iii) false claims related to the field of sport.

Within the broad themes found in this study, our analysis identified multiple sub-topics. The broad topic of 'politics', for example, is identified as comprising 10 sub-topics, ranging from 'the actions and views of politicians' to 'the practical process of elections' and 'the size, legality and organisation of political protests'. And the broad topic of 'health' is identified as covering 5 sub-topics ranging from 'access to health services', to 'the effectiveness and risks of particular health treatments'. In this particular study, most items in the dataset related to between two and four such themes or sub-topics.

As well as reflecting the news-related focus of the fact-checkers' claim selection process, described above, the mixture of topics identified also likely reflects, at least in part, the financial rewards that come to users who spread information on trending topics likely to draw partisan interest - regardless of whether or not those who create this content are motivated by the topic themselves.[70]

The most frequent topic in the dataset related to what could be broadly described as "politics", appearing in 71 (63.4%) of the 112 pieces of content examined. Content related to foreign countries featured in 43 (38.4%). Business and the economy, and content related to crime and justice featured in 28 (25%) and 27 (24.1%), respectively. The dataset included 20 pieces (17.9%) of content related to violent unrest and war, 18 (16%) related to communities and migration, 17 (15.2%) related to celebrities and public figures, 15 (13.4%) related to disasters and emergencies, 14 (12.5%) related to governance - particularly conspiracy theories - and 10 (8.9%) related to health and medicine. Other topics - from media to the environment - appeared less frequently in this sample. Within the 19 broad topics covered in the dataset, we identified 48 of the possible 88 sub-topics.

Most frequent among content related to "politics", appearing in 58 (51.8%) of 112 entries overall, was content that misrepresented the actions or views of high-profile politicians; most often using AI-generated video or audio impersonation of prominent political figures to fake supposed policy announcements, or fabricate false accounts of incidents that never happened in reality. Other sub-topics within the broad theme of politics included content related to 'politicians health or family circumstances' (4 times); 'the size of political protests' (3 times); 'public opinion and support for politicians or issues' (2 times); 'political appointments or resignations' (2 times); 'arrest or prosecution of politicians' (1 time) and 'government spending' (1 time).

[70] See for example this joint BBC and Bureau of Investigative Journalism investigation into who was behind a wave of anti-immigration AI-fakes. Mariana Spring, " Anti-immigration AI videos traced to overseas fakers, BBC finds", 15 May 2026, https://www.bbc.co.uk/news/articles/ckgpyn30dp3o

Within the politics-related content field, three particular themes, or narratives, appeared with more frequency than others. More than one in five of the 112 pieces of content examined[71] used fabricated content to promote claims feeding into broader negative narratives about the then prime minister and the governing Labour party. The two commonest narratives, appearing in 30 entries, were, on the one hand, supposed government announcements of measures such as fines or increased charges for use of vital public services, at a time of known public concern about the cost of living,[72] and, on the other hand, fabricated announcements by the same authorities of measures limiting personal freedoms. These included AI-generated content with politicians announcing measures such as fines for people for eating the 'wrong' sort of food and a limit on the number of flights members of the public may take per year.[73] Another common theme related to the policies of authorities - including the police - toward particular communities - Muslims and migrants - appearing 11 times.[74]

[71] Twenty-three of the 112 entries in the dataset related to supposed government actions or policies. Twenty-one related to the actions or views of then Prime Minister Sir Keir Starmer in particular (Thirteen of these related to both).Five entries related to government overreach into civil liberties.
[72] See, for example: (i) Sian Bayley, "No, a £15 'clean water levy' isn't being rolled out across the UK," *Full Fact*, 1 October 2025, https://fullfact.org/economy/government-15-clean-water-levy-false/; (ii) Sian Bayley, "The PM hasn't said state pension will be reduced in April 2026," *Full Fact*, 21 October 2025, https://fullfact.org/economy/government-reduction-state-pension-april-2026-false/; and (iii) Nasim Asl, "People in the UK will not be fined £750 for using their heating after 9pm", *Full Fact*, 28 November 2025, https://fullfact.org/politics/9pm-heating-750-fine/.
[73] See, for example: (i) Evie Townend, "'PM announcement' of two flight limit is fake", *Full Fact*, 25 September 2025, https://fullfact.org/politics/false-claims-starmer-announce-flight-quota/; (ii) Charlotte Green, "Keir Starmer hasn't just announced mass 'AI scanner' surveillance of all UK phones," *Full Fact*, 17 October 2025, https://fullfact.org/technology/phone-access-ai-scanners-false-video-claim/; and, (iii) Leo Benedictus, "The government is not going to track people's shopping and fine them for buying unhealthy food," *Full Fact*, 21 November 2025, https://fullfact.org/health/fake-healthy-basket-bill-grocery-monitoring/.
[74] See, for example: (i) Leo Benedictus, "An image of policemen kneeling in front of a group of Muslims is probably AI," *Full Fact*, 30 June 2025, https://fullfact.org/crime/police-kneeling-to-muslims-ai/. (ii) Charlotte Green, "Video of soldier shouting at police officer is AI-generated," *Full Fact*, 18 December 2025, https://fullfact.org/world/video-british-soldier-police-officer-shouting-free-speech-AI/.

Most frequent among content related to foreign countries was content related to specific 'events in foreign countries' (23 times),[75] followed by 'foreign governments' actions or views' (16 times)[76] and 'own governments' relationship with a foreign country' (4 times). Save either where (i) citizens in the home country have strong attachment to the foreign country or (ii) the false claim relates to foreign threats to the home country, the harm-risk model suggests such claims often have limited potential for effects at home.[77] However, the study covered a period during which a lot of public attention was paid to individuals and events, or supposed events, in the United States, in Ukraine, and, in the final weeks of the study period, in Iran, with potential for domestic impact. Studies such as Baum and Groeling (2010) suggest false information related to events in foreign countries is more likely to be accepted as true than false information on domestic issues because of the audience's more limited knowledge of events overseas.[78]

A common theme related to domestic news concerned matters related to the 'individual economy' and 'cost of living', such as AI-generated content falsely claiming to show news of an introduction of charges for popular services, increases to the cost of living or taxes (18 times). Our analysis also identified a frequent sub-topic of AI-generated content feeding into belief in one or more known false conspiracy theories (9 times).[79] Other topics appeared with lesser frequency.

## Audience & evidence of audience responses

To understand the potential of disinformation to cause effects, it is essential to consider not only the nature of the content but also the size, nature and responses of the audience.[80] For some consequences to occur, tens of thousands of individuals must see and act on the false information concerned. In other cases, it needs only one individual to act to cause a significant effect.[81] In some cases, consequences come through the misinformed actions of members of the public. In other cases, actions occur through the decisions of policymakers and leaders. As noted in the work used in developing the harm-risk model, false information published in academic journals, think-tank papers, traditional media, and in political settings has proven potential to influence policymaker decisions directly.[82]

[75] See, for example: (i) Hannah Smith, "Image showing US B-2 supposedly 'shot down' by Iran is fake", Full Fact, 11 March 2026, https://fullfact.org/conflict/fake-B-2-shot-down-image/.
(ii) Sian Bayley, "Video of Ukrainian flag on Statue of Liberty is fake," Full Fact, 14 March 2025, https://fullfact.org/online/ukrainian-flag-statue-of-liberty-fake/.; (iii) Sian Bayley, "Fake video of Emmanuel Macron claiming Donald Trump blocked his car in retaliation for recognising Palestine circulates online," Full Fact, 9 October 2025, https://fullfact.org/world/president-macron-video-motorcade-palestine-fake/

[76] See, for example: (i) Sian Bayley, "Another AI-generated video of Donald Trump criticising Keir Starmer circulates online," *Full Fac*t, 30 January 2025, https://fullfact.org/us/donald-trump-criticising-keir-starmer-ai-2/.(ii) Sian Bayley, "Picture of Mexican president wearing a 'Make America Mexicana Again' hat is fake," *Full Fact*, 7 February 2025, https://fullfact.org/online/picture-mexican-president-make-america-mexicana-again-hat-fake/.

[77] See, for example: Charlotte Green, "Video appearing to show Hurricane Melissa from above is AI", *Full Fact*, 29 October 2025, https://fullfact.org/environment/hurricane-melissa-video-from-above-ai/.

[78] Matthew Baum and Tim Groeling, "War Stories: the causes and consequences of public views of war," *Princeton University Press*, 2010.

[79] See, for example: (i) Sian Bayley, "AI-generated image of Jeffrey Epstein 'alive in Tel Aviv' circulates online," *Full Fact*, 16 February, 2026, https://fullfact.org/world/jeffrey-epstein-fake-image-tel-aviv/; (ii) Charlotte Green, "Keir Starmer hasn't just announced mass 'AI scanner' surveillance of all UK phones," *Full Fact*, 17 October 2025, https://fullfact.org/technology/phone-access-ai-scanners-false-video-claim/.

However, while some AI-generated videos and audio of the type seen in this study have potential to be seen as real by policymakers[83] - and the use of information from AI chatbots has been shown to contribute to poor decisions by senior UK decisionmakers[84] - the potential for consequences of most of the content in the dataset comes through its exposure to the public online, and that that audience is large and growing.[85]

With researchers' access to audience data restricted by the major open platforms, and limited means to assess the spread of information on encrypted platforms, the data on audience reach is inevitably incomplete and the audience numbers set out below are almost certainly an underestimate of the actual reach of the content in the dataset.

Altogether, the available data shows, the 112 pieces of content examined in the database reached an audience of tens of millions of Internet users, if not more. Most content was targeted at and reached audiences in the United Kingdom; other content was aimed at and reached audiences across the world. While some received quite limited audiences - in the low thousands of views online[86] - 22 of the 112 pieces of content hit one million or more views. The highest audience numbers publicly recorded was more than 11 million views for an AI-generated still image on X, purporting to show a political protest in Turkey.[87] One version, on X, of an AI-faked image purporting to show the late convicted sex offender Jeffrey Epstein living in Israel today, received more than 6 million views on that platform alone. Cross-platform spread was common.

---

[80] Chapter 4 ('Where and when false information has effect') in: Peter Cunliffe-Jones, "Fake news: What's the harm?" 2025, *University of Westminster Press*. https://doi.org/10.16997/mpub.14614695. pp 37-53.

[81] For a customer boycott to affect most major retail companies' bottom line, tens of thousands of people need to decide to boycott the company's products. See this from 2025. https://bpigroup.com/the-410m-cost-of-misinformation-for-coca-cola-rethinking-reputation-in-a-post-truth-world/. By contrast, it took one individual to see and act on the "Pizzagate" conspiracy theory about a supposed high-powered child sex ring in the United States. See: Matthew Haag and Maya Sa*lam,* "Gunman in 'Pizzagate' Shooting Is Sentenced to 4 Years in Prison," *The New York Times*, June 22, 2017, https://www.nytimes.com/2017/06/22/us/pizzagate-attack-sentence.html.

[82] Chapter 4 ('Where and when false information has effect') in: Peter Cunliffe-Jones, "Fake news: What's the harm?" 2025, *University of Westminster Press*. https://doi.org/10.16997/mpub.14614695. pp 37-53.

[83] Turkey Today, "AI video fools African leader into contacting Macron over fabricated French coup news", 17 December 2025, https://www.turkiyetoday.com/world/ai-video-fools-african-leader-into-contacting-macron-over-fabricated-french-coup-3211517

[84] Neha Gohil, "West Midlands police chief apologises after AI error used to justify Maccabi Tel Aviv ban," The Guardian, 14 January 2026, https://www.theguardian.com/uk-news/2026/jan/14/west-midlands-police-chief-apologises-ai-error-maccabi-tel-aviv-ban

[85] Felix Simon, Rasmus Kleis Nielsen & Richard Fletcher, "Generative AI and news report 2025: How people think about AI's role in journalism and society," *Reuters Institute for the Study of Journalism*, 7 Oct 2025 https://reutersinstitute.politics.ox.ac.uk/generative-ai-and-news-report-2025-how-people-think-about-ais-role-journalism-and-society

It is worth noting that - while the size of the audience does influence the potential for consequences in cases where large-scale action is needed to cause the effect - a large audience does not make consequences inevitable. Instead, it can be seen from the dataset that more than half of the pieces of content viewed one million times or more had no or limited potential for consequences for multiple reasons. These cases included entries which were found to be only narrowly inaccurate[88] related to topics on which any action an audience might take would not be consequential,[89] and entries where those who believed the claim had no capacity or motivation to act in a way that would cause a specific substantive consequence.[90] Audience size may make some consequences more likely, but not all.

## Formats of AI-generated disinformation

Of the 112 pieces of content assessed in the dataset, 66 were video-based and 46 were image-based. There was some variance within the category of videos in terms of the primary focus of the fact check. In 26 cases, the potential AI elements were in the video footage alone. In 21 cases, the potential AI elements were in the audio content of the video. In the remaining 19 cases, both the video and audio elements were involved.

[86] AI-generated content that spread a false claim that a named Conservative MP had decided to defect to Reform UK was discussed in the media, and in Prime Minister's Questions in parliament on 3 December 2025 but the content itself received only around 2,000 views online. Sian Bayley, "Video of George Freeman MP announcing defection to Reform UK is fake," Full Fact, 20 October 2025.
[87] Sian Bayley, "Image of fictional characters at the Turkey protests is AI-generated," Full Fact, 31 March 2025, https://fullfact.org/online/turkey-protests-fictional-characters-ai/.
[88]See, for example: Sian Bayley, "Viral image of Nicolás Maduro being led away from plane isn't real," *Full Fact*, 6 January 2026, https://fullfact.org/world/picture-maduro-seized-us-military-ai/; (ii) Sian Bayley, "Grok and Google Lens AI overviews claim fake imagery shows Huntingdon train attack," *Full Fact*, 5 November 2025, https://fullfact.org/crime/grok-google-lens-ai-imagery-train-attack.
[89] See, for example: (i) Charlotte Green, "Image of supercar surrounded by flames in Los Angeles is digital creation," *Full Fact*, 15 January 2025, https://fullfact.org/us/la-fires-supercar-image-digital-creation/; (ii) Sian Bayley, "Pictures of Rishi Sunak in a hospital bed are fake," *Full Fact*, 11 March 2026, https://fullfact.org/politics/rishi-sunak-hospital-pictures-fake/.
[90] See, for example: (i) Charlotte Green, "Video appearing to show Hurricane Melissa from above is AI", *Full Fact*, 29 October 2025, https://fullfact.org/environment/hurricane-melissa-video-from-above-ai/; (ii) Hannah Smith, "*Fake image of burning Burj Khalifa circulates online*", *Full Fact*, 2 March 2025, https://fullfact.org/world/burj-khalifa-ai-iran/

## Video with AI-generated visuals only

This category includes fully synthetic video clips, typically presenting fabricated news events, conflict scenes, or public figures in false scenarios. Examples include fake footage of Israeli strikes on Iran, a mass funeral in Sudan, and Los Angeles wildfire scenes. Audio elements were often present but the primary focus of the fact checks was on whether the videos depicted real events accurately.

## Videos with AI-generated audio only

The vast majority of videos in this category featured only incidental visuals, such as a collection of slides or muted real footage. While these could have included AI-generated visuals, the focus of the fact check was in the audio elements. In all cases, there was an apparently AI-generated audio overlay. Many of these examples were designed to mimic the voice and intonation of UK Prime Minister Keir Starmer.

A significant sub-group of this category includes fabricated announcements claiming that different personal freedoms of UK citizens are being taken away. This content included claims about airport toilet fees, Christmas decoration taxes, NHS charging for over-60s, and payslip deductions. The template is highly repeatable and potentially appears to be used by the same or related actors across multiple cases.

The main point of uncertainty in these checks was whether the voice mimicry was the result of AI-generation or an impersonator. Sometimes this was distinguishable because of vocal glitches which made AI-generation far more likely than a real person, but in other cases there were insufficient clues to be certain of this.

## Videos with AI-generated video and audio elements

Videos in this category most often featured real footage as a template for AI manipulation, coupled with apparently AI-generated audio. For example, a politician's lip movements are altered in a speech and new audio is overlaid over the top, synchronised with the lip movements. Common parlance for this form of content would be "lip-sync deepfakes".

## Images only

Static AI-generated images, often designed to look like photographs. Common use cases include fabricating meetings between political figures (Farage/Epstein, Trump/Epstein), and placing politicians in compromising or sentimental scenarios to shift public perception.

## Platform and actor types spreading AI-generated disinformation

A review of the dataset shows that AI-generated disinformation circulates on all major online platforms checked - Facebook, Instagram, Threads, TikTok, and X/Twitter. Many entries were found to have spread across multiple platforms, suggesting coordinated or at minimum copycat multi-platform seeding. With a relatively limited sample size, it was possible to draw only tentative conclusions about differences in behaviours on different platforms. Noting the sample size, the patterns suggested higher volumes of content identified on Facebook but greater virality on X/Twitter. They also indicated a focus on Facebook towards UK-targeted information and health treatment scams. X/Twitter accounted for more viral political content in general, while content on TikTok targeted younger audiences, including with health-related scams. Given both the sample size and a weighting in the factcheckers' selection process toward claims circulating on Meta-owned platforms, it is, however, not possible to confidently assert that these indications prove how much such information appears on different platforms. Further research on this point may be useful.

As is common with other studies of disinformation,[91] many of those who create and/or spread AI-generated false information do so openly - in public settings, traditional media, and from easily identified online accounts and platforms. In the dataset used in this study, it is possible to clearly identify: (i) known political actors and support groups; (ii) known public figures and organisations, (iii) openly labelled parody or satire online accounts and (iv) traceable online accounts using false information to promote scams, fake investment schemes, and supposed health products and services. In the former category, it was possible to identify a number of accounts that appeared pro-the Reform UK party creating and sharing large amounts of AI-generated disinformation - including both content about particular social issues and content impersonating the then prime Minister and mayor of London. In the second category it was also possible to identify AI-generated disinformation shared by prominent public figures in the United States, and elsewhere, such as X/Twitter owner Elon Musk, and a retired US general and former commander of US Special Operations Command. The latter shared an AI-altered image of the shooting of protestor Alex Pretti by US security forces in Minneapolis in January 2026. That post alone received more than 9 million views. However, where those who create and share false information wish to disguise or hide their identity, it is often challenging, even for platforms with access to privileged account-level information to identify those behind the content. Given the hidden or disguised identity of some of the accounts found sharing the politically-partisan disinformation noted in the earlier sections, it is likely that at least some such content was created by teams who create and share such content for the financial rewards it brings, rather than to achieve a political or social goal.[92]

[91] Peter Cunliffe-Jones, "Fake news: What's the harm?" 2025, *University of Westminster Press*. https://doi.org/10.16997/mpub.14614695 (pp 70-75).

[92] Mariana Spring, " Anti-immigration AI videos traced to overseas fakers, BBC finds", 15 May 2026, https://www.bbc.co.uk/news/articles/ckgpyn30dp3o

# 5. Harm Analysis

## Overview

**Analysis of the dataset shows that 94** (or 83.9%) of the 112 entries viewed by tens of millions of people between them **created a substantively false or misleading understanding among those who believe them.** These claims thus contributed, at an incremental level, to the broad disinformation effect of reducing trust, and/or reasons to trust, information circulating in public.[93] Studies show this broad disinformation effect[94] has potential for significant real world consequences in fields including public health[95] and support for democracy among others.[96]

**At the same time, our analysis showed that 46** (or 41.1%) of the 112 examples **met all the criteria in the model to cause, or to have caused or contributed to, specific, substantive consequences, or harms,** for individuals and society, with 66 (58.9%) of the entries showing no such potential. The risks of harm identified included potential to (i)

[93] While some false claims - in domestic politics or foreign affairs - are more salient than others and have greater impact than others on trust in information, the model used in this study identifies all substantively false or misleading information as contributing at some level to the public's reasons to distrust information in the public domain.
[94] For the impact that large-scale exposure has on public trust in information, see: Michael P  Lynch, "Fake News and the Internet Shell Game," The New York Times, November 28, 2016, https://www.nytimes.com/2016/11/28/opinion/fake-news-and-the-internet-shell-game.html.
[95] This 2020 study showed the perception of being disinformed reduced adherence to public health guidance: Michael Hameleers, Toni Van der Meer and Anna Brosius, "Feeling 'disinformed' reduces compliance with COVID-19 guidelines," *Harvard Misinformation Review*, May 31, 2020, https://misinforeview.hks.harvard.edu/article/feeling-disinformed-lowers-compliance-with-covid-19-guidelines-evidence-from-the-us-uk-netherlands-and-germany/.
[96] This 2024 study showed the perception that disinformation is widespread affects satisfaction with democracy: Andreas Jungherr & Adrian Rauchfleisch, "Negative downstream effects of alarmist disinformation discourse: Evidence from the United States" *Political Behavior*, 2024, 2123–2143. https://doi.org/10.1007/s11109-024-09911-3.

contribute to incidents of serious social unrest and vigilante violence,[97] (ii) justify or trigger abuse sufficiently serious to affect long-term mental and physical health,[98] (iii) cause direct financial harm to individuals via online scams,[99] (iv) cause direct harms to individuals' health by promoting ineffective, and potentially harmful, health supplements for women experiencing the menopause,[100] (v) contribute over time to a change in public

---

[97] For example, in February 2025, an AI-generated image purporting to show a "migrant mob" wielding axes in a hospital in Birmingham, was shared by Elon Musk and received more than 1.7 million views on social media. No such incident took place but the claim fed into a broad anti-migrant narrative. https://fullfact.org/online/birmingham-hospital-image-not-real-masked-men-axes/

Studies of the effects of (i) repetition on perception and (ii) 'trigger information' such as a sense of outsider threat - show such information has potential to contribute to circumstances in which violence occurs.

For the effects of repetition on perception see, for example, Floyd Allport and Milton Lepkin, "Wartime Rumors of Waste and Special Privilege: Why Some People Believe Them," *The Journal of Abnormal and Social Psychology*, 40.1, (1945), https://psycnet.apa.org/record/1945-01987-001. And Lisa Fazio, Raunak Pillai and Deep Patel, "The effects of repetition on belief in naturalistic settings," *Journal of Experimental Psychology*, 151.10, (2022): 2604-2613, https://pubmed.ncbi.nlm.nih.gov/35286116/.

For the effects of false claims of outsider threat see: Shakuntala Banaji, Ram Bhat, Anushi Agarwal, Nihal Passsanha and Mukti Pravin, "WhatsApp vigilantes: An Exploration of citizen reception and circulation of WhatsApp misinformation linked to mob violence in India," LSE, 2019, https://www.lse.ac.uk/media-and-communications/assets/documents/research/projects/WhatsApp-Misinformation-Report.pdf

[98] For example, in December 2025, an AI-generated image circulated online showing a person resembling Arsen Ostrovsky - a man injured in the Bondi Beach mass shooting - smiling as a woman applied fake blood to his face. https://fullfact.org/world/bondi-beach-bloodied-victim-fake-ai-image/. Audience responses to this post included antisemitic abuse, accusations of staging the incident, and threats levelled against Mr Ostrovksy.

Studies show that verbal and physical abuse at such a level have potential to harm the long-term mental and physical health of those targeted.

[99] For example, AI-generated content in an article in December 2025 claimed that financial expert Martin Lewis had been arrested for encouraging followers to use a crypto currency trading platform called Swiftgate Montark. https://fullfact.org/economy/fake-bbc-article-martin-lewis-arrested-facebook/

The article encouraged his followers to use the platform. Audience responses indicated some intended to do so, exposing them to potential financial loss and ID theft. Mr Lewis has spoken publicly about individuals defrauded by many such AI-generated scams using his faked image. See: https://www.thetimes.com/uk/crime/article/martin-lewis-meta-deepfake-fraud-victim-90hv5tsgr.

[100] For example, in December 2025, a video shared online presented an AI-generated content version of a real public health expert purportedly promoting use of supposed "health supplements" for women experiencing the menopause. https://fullfact.org/health/academics-deepfaked-tiktok-wellness-nest/.

Studies show the use of AI-generated fakes of known public health figures promoting untested, ineffective and sometimes harmful treatments is growing. See, for example: Chris Stokel-Walker, "Deepfakes and doctors: How people are being fooled by social media scams" BMJ 2024; 386. https://doi.org/10.1136/bmj.q1319.

Many studies show that members of the public presented with false claims about the supposed benefits of untested or ineffective health treatments will take up those treatments and stop use of effective treatment. See, for example: Nathan Thielman, Jan Ostermann, Kathryn Whetten, Rachel Itemba, Dafrosa Itemba, Venance Maro, Brian Pence and Elizabeth Reddy, "Reduced Adherence to Antiretroviral Therapy among HIV-infected Tanzanians Seeking Cure from the Loliondo Healer," *Journal of Acquired Immune Deficiency Syndrome*, 2014, https://doi.org/10.1097/01.qai.0000437619.23031.83.

trust in and engagement with the police,[101] (vi) contribute over time to a change in public attitudes to climate issues in ways that may impact public policy,[102] (vii) increase individuals' susceptibility to false conspiracy theories that have potential to cause specific harms,[103] and (viii) contribute over time to broader changes to public social and political attitudes over time in ways that may influence political and social events.[104],[105]

It is, of course, necessary to be clear about the scope and basis of these findings. The findings do not indicate that these outcomes will inevitably occur. Rather, the findings are that the three essential conditions identified in the harm risk model used[106] are in place for these outcomes to occur, or to have occurred. Further research would be required in

---

[101] For example, in June 2025, an AI-generated image with almost 160,000 views on X in the UK falsely claimed to show a group of police officers kneeling on the ground in front of a group of Muslims, criticising their supposed actions with a caption reading: "Only traitors bow to Invaders". https://fullfact.org/crime/police-kneeling-to-muslims-ai/.

Longitudinal studies of media effects on attitudes are rare, and complex, but - allied with multiple studies of the effects of repetition on perceptions - do support theories of long-term influence on attitudes.

See, for example: Florian Foos and Daniel Bischof, "Tabloid Media Campaigns and Public Opinion: Quasi-Experimental Evidence on Euroscepticism in England," *American Political Science Review*, 2021, 19-37, https://doi.org/10.1017/S000305542100085X

[102] For example, in September 2025, a false claim circulated that the UK government was planning to limit people to taking two flights per year - limiting people's freedoms in the name of action on climate. https://fullfact.org/politics/false-claims-starmer-announce-flight-quota/.

Multiple studies show that false claims about both the nature and causes of climate change and about the nature and effects of policies to counter its effects have influenced public attitudes around the world in ways that have potential to influence public policy.

See, for example, (i) Michael Tesler, "Elite Domination of Public Doubts About Climate Change (Not Evolution)," *Political Communication*, 35(2). (2017),306-326. https://doi.org/10.1080/10584609.2017.1380092. (ii) World Economic Forum, "3 charts that show how attitudes to climate science vary around the world," January 22, 2020, https://www.weforum.org/agenda/2020/01/climate-science-global-warming-most-sceptics-country/.

[103] See the example cited above which fed into a conspiracy theory about the government limiting people's individual freedoms for a variety of reasons.Studies of conspiracy-thinking show that believing in one false conspiracy theory makes it more likely that a person will believe in others; some of which have potential for harm.

[104] The dataset provides multiple examples of factually-false claims with potential to contribute to changes in attitude to social and political issues, where those or similar false messages are repeated over time.

Longitudinal studies of media effects are rare, and complex, but - allied with multiple studies of the effects of repetition on perceptions - do support theories of long-term influence on attitudes.

See, for example: Florian Foos and Daniel Bischof, "Tabloid Media Campaigns and Public Opinion: Quasi-Experimental Evidence on Euroscepticism in England," *American Political Science Review*, 2021, 19-37, https://doi.org/10.1017/S000305542100085X

[105] Beyond our sample, see this study of circulation in the UK of AI-generated videos strong with anti-immigration messages, researched by the BBC Bureau of Investigative Journalism. Mariana Spring, " Anti-immigration AI videos traced to overseas fakers, BBC finds", 15 May 2026, https://www.bbc.co.uk/news/articles/ckgpyn30dp3o

[106] (i) Evidence in the fact-check showed the claim created a substantively false understanding; (ii) evidence showed sufficient people believed the claim for it to have a particular effect; (iii) the evidence showed those who believed the claim had both capacity and potential motivation to act in a way that would cause the effect or could contribute to it if the claim or related claims were repeated over time.

such cases to prove beyond reasonable doubt whether such effects have in fact taken place.

On this basis, we discuss In the section below:

- the claims found to have no or limited potential for specific consequences, beyond a contribution to the broad effects on trust,
- the different fields and key examples of potential consequences identified
- the limited correlation between the virality, or scale of audience reached, and the perceived potential for effects
- the nature of informational effects: attitudinal & behavioural
- the nature of potential consequences - via cumulative and direct effects
- what can, and can't, be said of the scale, severity and duration of effects

## Claims that had no or limited potential for harm & why

Our analysis with the harms-risk assessment model found that 66 (58.9%) of the entries in our dataset had no or limited potential, either by direct or cumulative effect, for any specific, substantive consequences beyond their contribution to the broad disinformation effect on reliability and trust we already discussed.

To determine the potential of content to cause or contribute to consequences the model asks a series of questions in sequence. (i) Whether evidence shows the content was substantively false or misleading or only narrowly inaccurate; (ii) If the claim is found to be substantively false, whether any change in attitude or action this might cause would cause one or more of the real-world consequences identified in the model; (iii) If the above conditions are met, whether sufficient people have seen and believe the claim to cause or contribute to one or more of the consequences identified in the model that it might cause; and, (iv) If all the above conditions are met, the model asks whether those people who believe the claim have both the capacity and motivation to do cause or contribute to the specific consequence, or consequences, identified.

### Taking these questions in order, our analysis found that:

- Seventeen claims had no, or only limited, potential for misinformation effects - as only narrowly inaccurate

The model identifies claims as only narrowly inaccurate where the element that is inaccurate is either (i) immaterial to the audience's essential understanding of the situation or event or (ii) so marginally inaccurate as to make no meaningful difference to that essential understanding. This was the case for 17 entries in the dataset.

To illustrate this, one entry centred on a viral AI-generated image of former Venezuelan president Nicolás Maduro in detention by the United States in January 2026. The image was both artificial and inaccurate in details but was correct in the core understanding it caused or reinforced that the then president had been detained by US forces and brought to the United States.[107] Another entry focused on AI-generated images purporting to show scenes during a knife attack that took place inside a train in eastern England in November 2025. While the images were artificially created, evidence from verified eyewitnesses suggested they reflected the reality of what had happened. The false understanding related to the origins of the image, not the event itself; a narrowly inaccurate understanding.[108] In both these cases, and others, any action taken based on the understanding caused by the artificially-generated image would likely be indistinguishable from action taken based on authentic images. There was no, or only limited, potential for a specific, substantive misinformation effect.

- Nineteen remaining claims had no, or only limited, potential for specific consequences - as action based on the claim would not be consequential

Even if a claim is false, any change in attitude or actions it may cause may have no or limited potential to cause or contribute to one or more of the real-world consequences identified in the model, subject to its topic and context. This was the case for 19 entries in the database.
To illustrate this, consider the potential effects of an AI-generated image of a 'supercar' supposedly left intact amid bushfires around Los Angeles in January 2025. The understanding caused - that such a car had survived the fires that destroyed numerous homes and vehicles - was substantively false. However, the random nature of the claim that one particular car had survived the flames caused only a limited potential for those who believed it to act in a way that might cause any of the consequences named in the harm model.[109]

In another example, when a politician is running for office, claims about their state of health may have potential impact on voter intentions, as demonstrated in the US presidential election of 2024. In 2026, an AI-generated image of former Prime Minister Rishi Sunak supposedly recovering from an unnamed illness circulated online.[110] The claim was entirely false but - in this context - with Sunak secure in his parliamentary seat and no new election due - the actions or attitude the false claim might cause had no or limited potential to cause or contribute to any specific substantive consequence; only a false understanding or unnecessary "get well soon" card.

[107] Sian Bayley, "Viral image of Nicolás Maduro being led away from plane isn't real," *Full Fact*, 6 January 2026, https://fullfact.org/world/picture-maduro-seized-us-military-ai/;
[108] Sian Bayley, "Grok and Google Lens AI overviews claim fake imagery shows Huntingdon train attack," *Full Fact*, 5 November 2025, https://fullfact.org/crime/grok-google-lens-ai-imagery-train-attack/
[109] Charlotte Green, "Image of supercar surrounded by flames in Los Angeles is digital creation," *Full Fact*, 15 January 2025, https://fullfact.org/us/la-fires-supercar-image-digital-creation/
[110] Sian Bayley, "Pictures of Rishi Sunak in a hospital bed are fake," *Full Fact*, 11 March 2026, https://fullfact.org/politics/rishi-sunak-hospital-pictures-fake/.

In both these cases, and others, any action taken based on the understanding caused would likely be inconsequential.

- Twenty-four further claims had no, or only limited, potential for specific consequences - insufficient people believed the claim to cause the consequence

The model used in this study does not identify a false understanding as harmful in itself. Rather, it works on the premise that, for claims to have a substantive potential to cause specific real-world harm, a claim must cause or contribute to an action, or a change in attitude that leads to an action that causes harm. And, in most cases, that requires at least some of those who see the false claim to believe the claim to be true, or sufficiently true, or sufficiently likely to be true, for them to act on that basis.[111]

The model then considers the number of people who must believe in, and act on, the false claim in order to have a substantive potential to cause, or contribute over time, to a potential specific consequence.

Among many entries in the dataset that failed this test, for example, was an AI-generated video purporting to show a giant cruise ship dumping large amounts of human waste at sea. The video related to the harms done to the environment by such ships. To cause a substantive change in behaviour towards such companies requires that thousands believe and change behaviour based on the claim. However while the video was seen more than 1.8 million times, audience responses showed overwhelming scepticism, recognising the video as AI-generated "fake news", with no evidence of strong impacts on understanding or behaviour.[112]

- Six remaining claims had no, or only limited potential for specific consequences - as the audience didn't have capacity and motivation to cause consequence[113]

Finally, the model works on the premise that - even if sufficient people see and believe a false claim to cause or contribute to a particular effect - they must have the capacity and motivation to act in a way that would cause or contribute to the consequence in order for it to occur.

---

[111] See pages 152-155 in: Peter Cunliffe-Jones, "Fake news: What's the harm?" 2025, *University of Westminster Press*. https://doi.org/10.16997/mpub.14614695
[112] Charlotte Green, "Video showing cruise ship spilling waste into the sea isn't real," *Full Fact*, 25 August 2025, https://fullfact.org/environment/cruise-ship-sewage-sea-ai-generated/
[113] As these questions are asked sequentially, the model likely under-identifies the number of questions where audiences may have lacked capacity and motivation.

For example, individuals who falsely believe that a fake medication might cure them of a particular health condition must be able to acquire it, in the real world, in order for its false claims to have substantive potential for direct harm.[114] And for false information about a weather event to affect people's safety, they need to be in the line of that event.

Thus, for example, an entry in the dataset showed an AI-generated video purporting to show Hurricane Melissa which hit Jamaica in 2025. While audience responses showed many believed the video to be genuine, most of those who did so lived outside the track of the hurricane and did not have the capacity to act in a way that would have any substantive real world effects. Meanwhile, those who lived in Dubai and saw an AI-generated image purporting to show the Burj Khalifa ablaze after an attack by Iran knew the claim was false. And those outside the United Arab Emirates who believed the claim had no capacity to act on the false understanding in a way likely to cause a specific real-world consequence.[115]

## Eight different fields of potential consequences identified

**Our analysis of the dataset found that 46** (or 41.1%) of the 112 examples examined met all **the criteria to cause, or to have directly caused or contributed over time, to one or more specific, substantive consequences** identified in the risk model; a total of **78 potential consequences in eight broad fields** from causing or contributing to incidents of serious social unrest and/or individual vigilante violence to causing direct financial harm to individuals via online scams, hoaxes and causing harm to individual and public health.

### 1. Potential to cause, contribute to incidents of social unrest, vigilante violence

Analysis of our dataset showed eight AI-generated pieces of content - some viewed millions of times - made factually false claims of negative behaviour by - and preferential treatment by the authorities of - Muslims, or migrants,[116],[117] which our model suggests have a  similar substantive potential to contribute to negative public attitudes toward these communities. In this context, our harm risk model identified two examples of AI-generated disinformation in our dataset which had used AI-generated false content about Muslims' behaviour and the 'patriots' who combat them, which the model found had the potential to trigger or be used to justify incidents of unrest or violence. These were a widely shared AI-generated false claim of an attack on a hospital in Birmingham, causing both outrage and a sense of threat,[118] and an AI-generated image of a purported attack on migrant boats, which had potential to justify similar actions.[119]

---

[114] In 2026, claims about the supposed health benefits of unpasteurized milk spread widely in Hong Kong. The claims were false but widely believed. Unpasteurized milk is dangerous to health. However, the claims had little potential for harm as unpasteurized milk is not available in the city.
[115] Hannah Smith, "*Fake image of burning Burj Khalifa circulates online*", *Full Fact*, 2 March 2025, https://fullfact.org/world/burj-khalifa-ai-iran/

Incidents of social unrest and individual vigilante violence tend to result from multiple long-term and short-term factors. Social conditions change and societal tensions build over time for a variety of social, political and social-economic reasons.[120] When a trigger event leads to unrest or violence, this is often portrayed in media reports as "the cause" but in fact plays only one, albeit important, part.[121] While longitudinal studies of the role of media exposure in this process are both rare and complex to carry out, the ones that have been published - allied with multiple studies of the effects of repeated exposure to information on perception[122] - support theories of the potential of information to contribute to long-term attitudinal changes.[123] And studies of social unrest and acts of violence show that, when such tensions are allowed to build, information about a perceived outsider threat or reports of wrongdoing by individuals or groups, has potential to act as either a trigger, or perceived justification for incidents of social unrest and individual acts of vigilante violence.[124] This two-stage sequence of long-term and short-term trigger effects was perhaps most recently demonstrated by the effects contributing to the outbreak of riots in the UK in 2024.

---

[116] See for example:Sian Bayley, "Fake video of UK teacher telling children to bow and chant 'Allahu Akbar' circulates online", *Full Fact*, 17 November 2025. https://fullfact.org/education/fake-video-uk-teacher-islam-likely-ai/

[117] See, for example, Leo Benedictus, "An image of policemen kneeling in front of a group of Muslims is probably AI," *Full Fact*, 30 June 2025, https://fullfact.org/crime/police-kneeling-to-muslims-ai/.

[118] Evie Townend, "Image of masked men carrying axes is not a real photo of an incident at a Birmingham hospital", *Full Fact*, 19 February 2025, https://fullfact.org/online/birmingham-hospital-image-not-real-masked-men-axes/.

[119] Sian Bayley, "Fake pictures of 'British vigilantes' slashing small boats shared online," *Full Fact*, 23 January 2026, https://fullfact.org/immigration/fake-pictures-slashing-small-boats/.

[120] The 'father' of the so-called "limited effects" school of media effects theory was social scientist Paul Lazarsfeld. See: Paul Lazarsfeld, Bernard Berelson and Hazel Gaudet, "*The people's choice," Columbia University Press*, 1948

[121] This point is addressed in reports by DSIT reviewing the lead up to the 2024 Southport riots. See: https://committees.parliament.uk/work/8641/social-media-misinformation-and-harmful-algorithms/publications/.

[122] See, (i) one of the first studies of the effects on repetition on perceptions: Floyd Allport and Milton Lepkin, "Wartime Rumors of Waste and Special Privilege: Why Some People Believe Them," *The Journal of Abnormal and Social Psychology*, 40.1, (1945), https://psycnet.apa.org/record/1945-01987-001, and (ii) a more recent study of the effects of repetition on belief. Lisa Fazio, Raunak Pillai and Deep Patel, "The effects of repetition on belief in naturalistic settings," *Journal of Experimental Psychology*, 151.10, (2022): 2604-2613, https://pubmed.ncbi.nlm.nih.gov/35286116/.

[123] Florian Foos and Daniel Bischof, "Tabloid Media Campaigns and Public Opinion: Quasi-Experimental Evidence on Euroscepticism in England," *American Political Science Review*, 2021, 19-37, https://doi.org/10.1017/S000305542100085X

## 2. Potential to justify or trigger abuse affecting long-term health, how they live

False information has long been used to enact or encourage verbal or physical abuse of particular individuals or communities. The particular capacity of AI to impersonate individuals - either ridiculing or showing them acting in a negative light - adds a new and powerful means to encourage this sort of abuse. Some online abuse has limited impact beyond causing fleeting negative emotions in those targeted. However, the harms-risk model used in this study recognises that both online and offline abuse can sometimes reach such a level - with repeated and widespread insults and threats - that it affects the subject's long-term mental or physical health,[125] or creates such a threat that the victims are forced to seek protection or change where they live.[126] And recognising that effects are likely to be exacerbated when the abuse targets a particular individual, that individual has suffered prior traumas, threats are specific and detailed and appear realistic, and abuse is both widespread and repeated over time.

In the dataset for this study, our analysis identified eight examples of AI-generated content as having potential to contribute over time to creating or reinforcing attitudes that can give rise to incidents of verbal and/or physical abuse in certain circumstances. Our analysis identified one entry - an AI-generated image showing a person resembling Arsen Ostrovsky - a Jewish man injured in the Bondi Beach mass shooting of December 2025 - as having a substantive potential to have directly triggered sustained abuse of that one individual sufficient to have affected his physical or mental health over the medium or long-term.[127] There was no evidence it had such an effect.

---

[124] See: (i) the DSIT reports on the role of false information in the 2024 UK summer riots, https://committees.parliament.uk/work/8641/social-media-misinformation-and-harmful-algorithms/publications/, and (ii) this report on so-called "WhatsApp murders" in India in 2018.
Shakuntala Banaji, Ram Bhat, Anushi Agarwal, Nihal Passanha and Mukti Sadhana Pravin, "WhatsApp vigilantes: An Exploration of citizen reception and circulation of WhatsApp misinformation linked to mob violence in India," LSE, 2019, https://www.lse.ac.uk/media-and-communications/assets/documents/research/projects/WhatsApp-Misinformation-Report.pdf

[125] For evidence of the physiological effects of abuse, stress and fear, see: Karnatovskaia, LV, Johnson, MM, Varga, K, Highfield, JA, Wolfrom, BD, Philbrick, KL, Wesley Ely, E, Jackson, JC, Gajic, O, Ahmad, SR, Niven, AS. (2020) 'Stress and Fear: Clinical Implications for Providers and Patients (in the Time of COVID-19 and Beyond)'. Mayo Clinical Proceedings, 95(11). 2487-2498. https://doi.org/10.1016/j.mayocp.2020.08.028.

[126] Jamie Grierson, "Former police officer in hiding after being falsely linked to Henry Nowak arrest," *The Guardian*, 3 June 2026, https://www.theguardian.com/uk-news/2026/jun/03/former-officer-hampshire-hiding-after-being-falsely-linked-henry-nowak-arrest.

## 3. Cause direct financial harm to individuals via online scams

Online scams and cyber-enabled fraud were recently estimated to cost the UK economy around £14.7 billion annually, with individuals losing nearly £1.3 billion annually to payment fraud and scams.[128] While scams and frauds are not new, the particular ability of AI to impersonate individuals - from publicly-known figures shown endorsing products or services they would not support, to friends or family of a private individual falsely shown as being in danger and in need of funds - adds a new means for scammers to use in such frauds.

In the dataset for this study, our analysis identified AI-generated content claiming that financial expert Martin Lewis had been arrested for encouraging followers to use a crypto currency trading platform called Swiftgate Montark as having potential to defraud individuals who place trust in Lewis.[129] Numerous media and official reports show that such scams and frauds - using AI-generated fake endorsements by public figures or impersonation of people known to the victims are both widespread and growing in number.[130]

## 4. Cause direct harms to individual and public health

Numerous studies exist of different ways in which false information has potential to cause or contribute to harm to individual and public health.[131]

These include creating, or contributing to, false understandings of (i) prevalence, nature, causes & treatment of mental ill-health (e.g. a false claim misrepresenting the susceptibility of a particular community to mental illness); (ii) incidents and prevalence of suicide, self-harm (e.g. a false claim about a prominent individual ending their life by suicide);[132] (iii) access to, cost of access and quality of health services (e.g. a false claim about the cost of access to health services); (iv) the effectiveness and/or risks of different health treatments (e.g. promoting use of a harmful treatment or deterring use of an effective medication);[133] (v) the causes, effects, symptoms, means of spread, prevalence and/or susceptibility to health conditions (e.g. factors that contribute to incidence of a type of cancer or the means of transmission of a communicable disease); (vi) the effects on health of diet, activities and/or non-medicinal products; (vii) the actions and effects of health practitioners, drugs companies and health authorities (e.g. the actions of health workers spreading disease intentionally);[134] (viii) the actions of the public in relation to and/or affecting public health (e.g. a false claim about public adherence with health guidelines).

[127] Charlotte Green, “Fake AI image of Bondi Beach victim having blood applied circulates online,” Full Fact, 17 December 2025, https://fullfact.org/world/bondi-beach-bloodied-victim-fake-ai-image/
[128] Online Information Advisory Committee “Understanding Online Financial Harm Fraud, Scams and Disinformation Exposure Across Demographic Groups in the UK”, *Ofcom*, 27 November 2025 https://www.ofcom.org.uk/siteassets/resources/documents/about-ofcom/structure-and-leadership/online-information-advisory-committee/understanding-online-financial-harm-fraud.pdf?v=408231.
[129] Sian Bayley, “Fake ‘BBC’ article falsely claims that Martin Lewis has been arrested,” Full Fact, 19 December 2025, https://fullfact.org/economy/fake-bbc-article-martin-lewis-arrested-facebook/
[130] See: “What Martin Lewis told a Facebook fraud victim who lost £50,000,” *The Times,* June 2026. https://www.thetimes.com/uk/crime/article/martin-lewis-meta-deepfake-fraud-victim-90hv5tsgr.
[131] Israel Junior Borges do Nascimento, Ana Beatriz Pizarro, Jussara M Almeida, Natsasha Azzopardi-Muscat, Marcos Andre Goncalves, Maria Björklund and David Novillo-Ortiz, ”Infodemics and health misinformation: a systematic review of reviews”. *Bulletin of the World Health Organization*. 100(9). (2022). 544-561. https://doi.org/10.2471/BLT.21.287654.

In the dataset for this study, our analysis identified AI-generated content in a video purportedly showing a public health expert promoting use of ineffective and potentially harmful supposed "health supplements" for women experiencing the menopause.[135] Many studies show that when presented with false claims about the supposed benefits of untested or ineffective health treatments, many members of the public will take up those treatments and stop use of effective treatment,[136] and our harms-risk analysis model identified this AI-generated content as having substantive potential to harm the health of those who saw and responded. This entry is one of numerous such examples of AI-generated content showing known public health figures promoting untested, ineffective and sometimes harmful treatments - potentially harming public health.[137]

---

[132] See: "The contagion of suicidal behavior", *NCBI*, https://www.ncbi.nlm.nih.gov/books/NBK207262/
[133] See: (i) Pride Chigwedere, George Seage, Sofia Gruskin, Tun-Hou Lee and M Essex, "Estimating the lost benefits of antiretroviral drug use in South Africa," *Journal of Acquired Immune Deficiency Syndrome*, 49(4) (2008), 410-415, https://doi.org/10.1097/qai.0b013e31818a6cd5. (ii) Kelsey Piper, "The staggering death toll of scientific lies" Vox, August 26, 2024, https://www.vox.com/future-perfect/368350/scientific-research-fraud-crime-jail-time; (iii) Hamidreza Aghababaeian, Lara Hamdanieh and Abbas Ostadtaghizadeh, "Alcohol intake in an attempt to fight COVID-19: A medical myth in Iran," *Alcohol*, 88 (2022) 29-32, https://doi.org/10.1016/j.alcohol.2020.07.006.
[134] Deutsche Welle, "Ebola burial team attacked," September 20 2014, https://www.dw.com/en/ebola-burial-team-attacked-in-sierra-leone/a-17937340#.
[135] Leo Benedictus. "Revealed: how academics are being deepfaked on TikTok and Instagram to promote supplements," Full Fact, 5 December 2025, https://fullfact.org/health/academics-deepfaked-tiktok-wellness-nest/.
[136] Nathan Thielman, Jan Ostermann, Kathryn Whetten, Rachel Itemba, Dafrosa Itemba, Venance Maro, Brian Pence and Elizabeth Reddy, "Reduced Adherence to Antiretroviral Therapy among HIV-infected Tanzanians Seeking Cure from the Loliondo Healer," *Journal of Acquired Immune Deficiency Syndrome*, 2014, https://doi.org/10.1097/01.qai.0000437619.23031.83;
[137] Chris Stokel-Walker, "Deepfakes and doctors: How people are being fooled by social media scams" BMJ 2024; 386. https://doi.org/10.1136/bmj.q1319.

## 5. Contribute to reduced trust in, engagement with the police, justice systems

Research has identified many ways that misinformation and disinformation affect public trust in, and engagement with, and outcomes of the police and justice systems. These range from the way misinformation affects the outcome of particular court cases and how the justice system functions[138] to how false information affects public trust in and engagement with the police and courts system, including through the longitudinal effect of such information on public attitudes.[139]

In the dataset for this study, our analysis identified three entries as having a substantive potential to contribute to reduction in trust in, and engagement with, the police. Two viral AI-generated images purportedly showed police officers kneeling in front of a group of Muslims,[140] and an AI-generated video spread online supposedly showed a soldier berating a policeman for failing to counter crime by "invaders".[141]

## 6. Contribute to change in public attitudes in ways that may impact climate policy

Many years of research show that false narratives about both the nature and causes of climate change and the policies to counter its effects, have influenced public attitudes to climate action in ways that have influenced public policy in countries around the world.[142] In recent years, as more and more evidence has emerged of the reality of climate change, disinformation has shifted to exaggerating the impact of climate action on people's lives.

---

[138] See (i) Steven Frenda, Rebecca and Elizabeth F. Loftus, "Current Issues and Advances in Misinformation Research," *Current Directions in Psychological Science,* 20.1, (2021). https://doi.org/10.1177/0963721410396620; (ii) Elizabeth F. Loftus, "Planting misinformation in the human mind. A 30-year investigation of the malleability of memory," *Learning and memory*, 12, (2005), 361-366, https://doi.org/10.1101/lm.94705.

[139] Florian Foos and Daniel Bischof, "Tabloid Media Campaigns and Public Opinion: Quasi-Experimental Evidence on Euroscepticism in England," *American Political Science Review*, 2021, 19-37, https://doi.org/10.1017/S000305542100085X

[140] Leo Benedictus, "An image of policemen kneeling in front of a group of Muslims is probably AI," *Full Fact*, 30 June 2025, https://fullfact.org/crime/police-kneeling-to-muslims-ai/.

[141] Charlotte Green, "Video of soldier shouting at police officer is AI-generated," *Full Fact*, 18 December 2025, https://fullfact.org/world/video-british-soldier-police-officer-shouting-free-speech-AI/

[142] See: (i) James Hoggan and Richard Littlemore, "Climate cover-up: The crusade to deny global warming," *Greystone Books*, 2009, 61-72; (ii) Naomi Oreskes and Erik M Conway, "Merchants of Doubt. How a handful of scientists obscured the truth on issues from tobacco smoke to global warming", *Bloomsbury Press*, 2010. (iii) Michael Tesler, "Elite Domination of Public Doubts About Climate Change (Not Evolution)," *Political Communication*, 2017,306-326. https://doi.org/10.1080/10584609.2017.1380092. (iv) World Economic Forum, "3 charts that show how attitudes to climate science vary around the world," January 22, 2020, https://www.weforum.org/agenda/2020/01/climate-science-global-warming-most-sceptics-country/.

In the dataset for this study, we identified one entry that used AI-generated content to suggest that the UK government was introducing a ban on people taking more than two flights a year.[143] Studies show that repeated exposure to this sort of false information[144] have potential to contribute to changes in public attitudes to measures to tackle climate change in ways that may impact public policy.

## 7. Increase susceptibility to conspiracy theories with potential to cause harms

Research shows a range of factors increase susceptibility to conspiracy beliefs, from life experiences to individual characteristics such as cognitive-perceptual schizotypy (which involves jumping to conclusions and bias against disconfirmatory evidence) and sub-clinical psychopathy.[145]

Studies of conspiracy beliefs also suggest that belief in one conspiracy theory does not simply show propensity to believe in conspiracy theories but makes other conspiracy theories more believable. The 'fact' that a conspiracy exists in one domain is seen, by the conspiracist, as evidence of a pattern supporting their theories about other events. "Your feelings about each individual conspiracy theory are determined in large part by the extent to which you buy into an overarching set of all-purpose assumptions about how the world works," as one study noted.[146]

In the dataset for this study, our analysis identified 11 entries - some viewed millions of times - spreading AI-generated content feeding into different conspiracy theories about the government impingeing on people's privacy[147] and personal freedoms[148] in ways that had substantive potential to influence political and social attitudes and promote conspiracy beliefs with the potential to contribute to harm.

---

[143] Evie Townend, "'PM announcement' of two flight limit is fake", *Full Fact*, 25 September 2025, https://fullfact.org/politics/false-claims-starmer-announce-flight-quota/

[144] See, (i) one of the first studies of the effects on repetition on perceptions: Floyd Allport and Milton Lepkin, "Wartime Rumors of Waste and Special Privilege: Why Some People Believe Them," *The Journal of Abnormal and Social Psychology*, 40.1, (1945), https://psycnet.apa.org/record/1945-01987-001, and (ii) a more recent study of the effects of repetition on belief. Lisa Fazio, Raunak Pillai and Deep Patel, "The effects of repetition on belief in naturalistic settings," *Journal of Experimental Psychology*, 151.10, (2022): 2604-2613, https://pubmed.ncbi.nlm.nih.gov/35286116/.

[145] Tom Buchanan and James Kempley, "Individual differences in sharing false political information on social media: Direct and indirect effects of cognitive-perceptual schizotypy and psychopathy," *Personality and Individual Differences*, 182 (2021), https://doi.org/10.1016/j.paid.2021.111071.

[146] Rob Brotherton, "Suspicious minds - Why we believe conspiracy theories," *Bloomsbury Sigma*, 2015. (pp 96-98)

## 8. Contribute to changes to social, political attitudes in ways that could affect social, political outcomes

The most fiercely argued debate over the effect of misinformation relates to effects on political and social attitudes and behaviour. As far back as 17th century philosopher Francis Bacon,[149] thinkers, writers, and social and cognitive scientists have argued that while individuals' factual understandings will often change in the face of new evidence, people's views and opinions, once formed, tend to be slow to change.[150] And, where attitudes do change, a wide range of social and societal factors - not only media - play a key role.[151] At the same time, multiple studies have, over many years, shown the powerful effects of repetition on audience perception.[152] And more recent longitudinal studies of media effects on attitudes do support theories of long-term influence on both attitudes and behaviour.[153]

---

[147] See: (i) Charlotte Green, "Keir Starmer hasn't just announced mass 'AI scanner' surveillance of all UK phones", Full Fact, 17 October 2025, https://fullfact.org/technology/phone-access-ai-scanners-false-video-claim/; (ii) Leo Benedictus, "The government is not going to track people's shopping and fine them for buying unhealthy food", Full Fact, 21 November 2025, https://fullfact.org/health/fake-healthy-basket-bill-grocery-monitoring/.

[148] See: (i) Evie Townend, "'PM announcement' of two flight limit is fake", *Full Fact*, 25 September 2025, https://fullfact.org/politics/false-claims-starmer-announce-flight-quota/; (ii)Leo Benedictus, "The government is not going to track people's shopping and fine them for buying unhealthy food", Full Fact, 21 November 2025, https://fullfact.org/health/fake-healthy-basket-bill-grocery-monitoring/.

[149] Francis Bacon, "*Novum Organum",* 1620, https://oll.libertyfund.org/title/bacon-novum-organum.

[150] Brendan Nyhan and Jason Reifler, "When corrections fail: The persistence of political misperceptions", *Political Behavior*, 32(2), 303–330 (2010). https://doi.org:10.1007/s11109-010-9112-2.

[151] Paul Lazarsfeld, Bernard Berelson and Hazel Gaudet, "*The people's choice," Columbia University Press*, 1948

[152] See: (i) Floyd Allport and Milton Lepkin, "Wartime Rumors of Waste and Special Privilege: Why Some People Believe Them," *The Journal of Abnormal and Social Psychology*, 40.1, (1945), https://psycnet.apa.org/record/1945-01987-001. (ii) Lisa Fazio, Raunak Pillai and Deep Patel, "The effects of repetition on belief in naturalistic settings," *Journal of Experimental Psychology*, 151.10, (2022): 2604-2613, https://pubmed.ncbi.nlm.nih.gov/35286116/. (iii) Doris Lacassagne, Jeremy Bena, Olivier Corneille, "Is Earth a perfect square? Repetition increases the perceived truth of highly implausible statements," *Cognition*, Vol 223, (2022), https://doi.org/10.1016/j.cognition.2022.105052.

[153] Florian Foos and Daniel Bischof, "Tabloid Media Campaigns and Public Opinion: Quasi-Experimental Evidence on Euroscepticism in England," *American Political Science Review*, 2021, 19-37, https://doi.org/10.1017/S000305542100085X

In the dataset for this study, our analysis identified AI-generated content related to broadly "political" themes in 71 (63.4%) of the 112 entries. Most frequent, appearing in 58 (51.8%) of the 112 entries, was content that misrepresented the actions or views of high-profile politicians; most often using AI-generated video or audio impersonation of prominent political figures to fake supposed policy announcements, or fabricate false accounts of incidents that never happened in reality. 18 entries (16% of the total) misrepresented the behaviour, treatment of or attitudes to particular communities and 14 (12.5%) created a false understanding of issues of governance - of which 11 related to conspiracy theories.

Of these entries that misrepresented the actions or views of politicians, the most frequently targeted figure was the prime minister in place during the study period, Sir Keir Starmer. The two commonest narratives were AI-generated impersonations of Starmer announcing unpopular measures such as supposed fines, fees for accessing toilets, taxes on Christmas decorations, or increased charges for use vital public services,[154] or measures limiting personal freedoms such as limits on cash withdrawals or a ban on carrying hot drinks in public. Another common political theme related to the actions of the police toward particular communities - Muslims and migrants[155] The entries featuring AI-generated versions of Starmer making announcements appeared to use a near-identical structure and target audiences already known or primed to distrust the government. The repeated nature of the template suggests an organised or at minimum copycat operation, which the current dataset does not have the fields to identify or characterise. It is not possible, within the limits of this study, to assess the extent to which such factually false information affected public attitudes and behaviour in this period, but our findings show that the conditions to do so were met.

## Limits of correlation between virality & potential for effects

The relationship between the size of audience a false claim reaches and its potential for effects is a more complex one than often thought.

---

[154] See, for example: (i) Sian Bayley, "No, a £15 'clean water levy' isn't being rolled out across the UK," *Full Fact*, 1 October 2025, https://fullfact.org/economy/government-15-clean-water-levy-false/; (ii) Sian Bayley, "The PM hasn't said state pension will be reduced in April 2026," *Full Fact*, 21 October 2025, https://fullfact.org/economy/government-reduction-state-pension-april-2026-false/; and (iii) Nasim Asl, "People in the UK will not be fined £750 for using their heating after 9pm", *Full Fact*, 28 November 2025, https://fullfact.org/politics/9pm-heating-750-fine/.

[155] See, for example: (i) Leo Benedictus, "An image of policemen kneeling in front of a group of Muslims is probably AI," *Full Fact*, 30 June 2025, https://fullfact.org/crime/police-kneeling-to-muslims-ai/. (ii) Charlotte Green, "Video of soldier shouting at police officer is AI-generated," *Full Fact*, 18 December 2025, https://fullfact.org/world/video-british-soldier-police-officer-shouting-free-speech-AI/.

Where consequences occur due either to (i) decisions by a 'one in a million' individual,[156] or (ii) the collective effect of decisions of tens or hundreds of thousands of people,[157] claims need to reach large audiences for the consequences to be likely to occur. However in other cases, consequences may occur due to a decision by one very specific individual, reached not by spreading the information to a large audience but through more focused means. In 2000, for example, very few South Africans knew of a false theory published in an academic journal by German-American molecular biologist Peter Duesberg that HIV was not the cause of AIDS. One of those who was made aware was South African President Thabo Mbeki who changed the country's health public policy as a result, causing disastrous outcomes for public health in South Africa as a consequence.[158]

The question in such cases is not how many people are reached by the claim but does the claim reach the specific person or people able to act in a way that will cause the consequence.

Furthermore, our study shows again that even where a false claim does go viral, it may do nothing more than cause a widespread false understanding. In the dataset used for this study, our analysis found that a clear majority of viral entries - viewed one million times or more - had no or limited potential for consequences, the virality of the claim proving no guarantee of impact. This included entries found to be only narrowly inaccurate[159] related to viral but inconsequential topics,[160] and entries where those who believed the claim had no capacity or motivation to act in a way that would cause a specific substantive consequence.[161]

---

[156] In 2016, hundreds of thousands of Americans saw and believed the "Pizzagate" conspiracy theory about a supposed high-profile child sex ring operating from a restaurant in Washington DC. Most who did were not motivated to act on that claim by attacking the restaurant concerned but one was. The fact the claim reached so many people increased the potential of an attack. See: Matthew Haag and Maya Sa*lam,* "Gunman in 'Pizzagate' Shooting Is Sentenced to 4 Years in Prison," *The New York Times*, June 22, 2017,

[157] For example, in most cases, it requires the decision of tens or hundreds of thousands of people to change how they will vote, in order for political misinformation to affect the outcome of a major election.

[158] Pride Chigwedere, George Seage, Sofia Gruskin, Tun-Hou Lee and M Essex, "Estimating the lost benefits of antiretroviral drug use in South Africa," *Journal of Acquired Immune Deficiency Syndrome*, 2008, 410-415, https://doi.org/10.1097/qai.0b013e31818a6cd5.

[159] See, for example: Sian Bayley, "Viral image of Nicolás Maduro being led away from plane isn't real," *Full Fact*, 6 January 2026, https://fullfact.org/world/picture-maduro-seized-us-military-ai/; (ii) Sian Bayley, "Grok and Google Lens AI overviews claim fake imagery shows Huntingdon train attack," *Full Fact*, 5 November 2025, https://fullfact.org/crime/grok-google-lens-ai-imagery-train-attack.

[160] See, for example: (i) Charlotte Green, "Image of supercar surrounded by flames in Los Angeles is digital creation," *Full Fact*, 15 January 2025, https://fullfact.org/us/la-fires-supercar-image-digital-creation/; (ii) Sian Bayley, "Pictures of Rishi Sunak in a hospital bed are fake," *Full Fact*, 11 March 2026, https://fullfact.org/politics/rishi-sunak-hospital-pictures-fake/.

[161] See, for example: (i) Charlotte Green, "Video appearing to show Hurricane Melissa from above is AI", *Full Fact*, 29 October 2025,

## Scale of effects

The scale, severity and duration of consequences varies considerably depending on the topic and context of the claim. In the dataset used in this study, our analysis found that - if consequences were to have occurred - one of the outcomes would have affected a specific individual targeted by abuse. Eight of the outcomes would affect, or would have affected individuals in general - for example scams that cause individuals to suffer a financial loss or false claims about health treatments that are taken up by individuals not part of any specific group. Ten of the outcomes, if they were to occur, would affect or would have affected members of a specific group or community - for example abuse of Muslims or trans people. And 39 would affect or would have affected the whole or a large part of society - for example via effects on political or social attitudes with the potential to affect political or social outcomes. In 20 cases, where the potential outcomes involved things such as the effects of conspiracy theory beliefs, it was not possible to determine the scale of population who would be affected.

# 6. Recommendations

## Recommendations for social media and search platforms

- Social media and search platform disclosure policies should cover — at minimum — AI-generated or edited content belonging to the following categories:
    - The actions and views of politicians
    - The practical process of elections
    - The size, legality and organisation of political protests
    - The prevalence or causes of mental illness, as seen in other studies
    - Content purporting to show the sexual behaviour of private and public individuals
    - Symptoms of illnesses and the effectiveness and risks of particular health treatments
    - Claims related to sexuality, reproductive and sexual health
    - Claims related to the field of sport
    - Access to health services

---

https://fullfact.org/environment/hurricane-melissa-video-from-above-ai/; (ii) Hannah Smith, "*Fake image of burning Burj Khalifa circulates online*", *Full Fact*, 2 March 2025, https://fullfact.org/world/burj-khalifa-ai-iran/

- Accidents, disasters and emergencies
    - Violent unrest, conflict and war
    - Business and economy
    - Crime, justice, law and policing
    - Communities and migration
    - Costs and powers of the state, abuses of power and corruption
- Content creators who repeatedly fail to disclose AI generated content as directed by such community guidelines should be permanently barred from services on a three strikes basis.
- Work with other companies to develop technology to detect seeding and copycat content, to enable labelling of coordinated campaigns during periods of vulnerability like crises and elections.
- Participate in international standards for indirect disclosure techniques.
- Provide researcher access and enable transparency about the accuracy and reliability of detection tools used to moderate content and enforce policies.

## Recommendations for Government

- Champion phone and camera companies building content credentials into camera technology.
- Require generative AI companies offering services to UK users to apply automatic digital watermarking when AI-generated content is created.
- Require large online platforms, search engines and widely deployed AI systems to implement clear, consistent and interoperable standards for signalling the provenance of AI-generated content at the point of use (through labelling), building on emerging technical standards like SynthID and the C2PA specification.
- Introduce a duty on the largest online platforms and search engines to assess and mitigate any actual or foreseeable negative effects that their services pose to civic discourse, electoral processes and public security and public health.
- The government should establish a robust framework for independent researcher access to platform, search engine and AI system data, building on proposals from Ofcom and international best practice. This framework should enable secure, privacy-preserving and proportionate access for independent researchers, including accredited fact checkers, with provision for timely access during elections and other high-risk periods.

# 7. Conclusion

The potential risks posed by AI-generated disinformation and misinformation for individuals and society are growing in the United Kingdom. A substantive risk of negative consequences from false information examined in this study was identified in eight fields ranging from contributions to incidents of serious social unrest and vigilante violence, to harms to health and financial loss caused by online scams and fraud.

This report – using a harm-risk assessment model developed in a four-year research project published by University of Westminster Press in 2025[162] – analysed evidence from a dataset of 112 pieces of AI-generated or AI-altered disinformation or misinformation seen tens of millions of times across the UK between 1 January 2025 and 31 March 2026. Our analysis identified:

- 94 of the 112 entries (83.9%) as creating a substantively false or misleading understanding, i.e. a broad disinformation effect
- 18 of the entries (16.1%) as creating an understanding that was only narrowly inaccurate, i.e. no such effect

Beyond this broad disinformation effect, our analysis identified:

- 46 of the 112 entries (41.1%) as having a substantive potential to cause or have caused or contributed to, one or more of the specific substantive consequences set out in the model
- 66 of the entries (58.9%) were found to have either no or very limited potential for consequences.

These potential consequences were identified in eight broad fields, either directly, via public exposure to that specific piece of false information, or contributing by cumulative effect to:

> (1) Incidents of serious social unrest and/or individual vigilante violence; (2) Incidents of abuse sufficiently serious to affect individuals' health, behaviour; (3) Direct financial harm to individuals via online scams, hoaxes; (4) Harm to individual and public health; (5) Public engagement with the police, justice system; (6) Public attitudes to climate issues that impact public policy; (7) Susceptibility to false conspiracy theories with potential to cause harms; (8) Public social and political attitudes that influence political, social events.

[162] See: (i) Peter Cunliffe-Jones, "Fake news: What's the harm?" 2025, *University of Westminster Press*. https://doi.org/10.16997/mpub.14614695; (ii) Peter Cunliffe-Jones, "Fact checking what matters: How a harms-based model for selecting claims works", *Harvard Kennedy School (HKS) Misinformation Review,* July 2026. https://doi.org/10.37016/mr-2020-197.

Our analysis found that - *if consequences were to occur* - one of the outcomes would affect a specific individual targeted by abuse. Eight of the outcomes would affect individuals not specific to a particular group - for example scams that cause a financial loss and false claims about health treatments that harm the health of individuals. Ten of the outcomes would affect members of a specific group or community - for example abuse of Jews, Muslims, members of the police or trans people. And 39 of the outcomes would affect a large part, or the whole of society, for example via incidents of social unrest, effects on political or social attitudes or effects on climate mitigation policies.

Given the role that disinformation and misinformation has played in events in the United Kingdom in recent years – from financial loss caused to individuals by online scams,[163] to public health during the Covid-19 Pandemic,[164] to the post-Southport riots[165] – and the context facing the United Kingdom in the coming years, we believe these risks to be serious and growing.

At the same time, it is worth reiterating that our analysis found that 66 of the entries (58.9%) were found to have either no or very limited potential for specific, substantive consequences.

The potential of some AI-generated disinformation and misinformation to cause or contribute to serious harm, and for other such information to have limited potential for impact, reinforces the need for robust methodologies for distinguishing the potentially harmful and the inconsequential.

---

[163] Online Information Advisory Committee “Understanding Online Financial Harm Fraud, Scams and Disinformation Exposure Across Demographic Groups in the UK”, *Ofcom*, 27 November 2025 https://www.ofcom.org.uk/siteassets/resources/documents/about-ofcom/structure-and-leadership/online-information-advisory-committee/understanding-online-financial-harm-fraud.pdf?v=408231.

[164] See this study “The impact of misinformation on the COVID-19 pandemic” https://pmc.ncbi.nlm.nih.gov/articles/PMC9114791/

[165] See this DSIT committee report: https://committees.parliament.uk/work/8641/social-media-misinformation-and-harmful-algorithms/publications/.

# Glossary

## Key terms

- **'Substantive consequences, or harms':** "objectively verifiable substantive injuries to the interests of an individual, group, or wider society"
- **'Cause effects'**: "Almost all behaviour occurs as a result of a combination of factors. The model identifies information as causing a consequence where cited evidence shows the claim caused a false factual understanding and this understanding was an essential, if not sole, factor in the outcome that followed.
- **'Contribute to effects'**: "The model identifies false information as contributing, substantively, to a consequence where evidence shows it reinforces, at a substantive level, a false understanding which, in time, is essential to the behavioural effect.
- **'Caused beyond reasonable doubt'**: The model identifies content as having caused or contributed to an effect 'beyond reasonable doubt' where, after investigation no other reasonable explanation plausibly explains the behaviour.
- **'Substantive potential'**: The model identifies content as having 'substantive potential' to cause or contribute to the effect where (i) existing research evidence shows that such information has in the past caused or contributed to such an effect, in such a context, (ii) the content meets the criteria set out in the model for factors required to cause the effect.